\documentclass[preprint,aps,amsmath,superscriptaddress,tightenlines,nofootinbib]{revtex4} 
\usepackage{graphicx,axodraw} 
\usepackage{bm} 

\usepackage{epsfig}

\def\Dslash{D\!\!\!\!\slash}

\def\nslash{n\!\!\!\slash} 
 
\def\bnslash{\bar n\!\!\!\slash}

\def\epsslash{\varepsilon\!\!\!\slash}

\def\OMIT#1{}

 \newcommand{\nn}{\nonumber} 
 
\newcommand{\bn}{{\bar n}} 
\newcommand{\bea}{\begin{eqnarray}} 
\newcommand{\eea}{\end{eqnarray}} 
\newcommand{\beq}{\begin{equation}} 
\newcommand{\eeq}{\end{equation}}

\newcommand{\mcdot}{\!\cdot\!} 
 
\newcommand{\cD}{{\cal D}}

\begin{document} 

\preprint{ 
\begin{tabular}{r} 
ZU-TH 07/03 \\ 
\end{tabular} 
} 
 
\title{\phantom{x}\vspace{0.5cm} 
Subleading collinear operators and their matrix elements 
\vspace{0.5cm} }

\author{\mbox{Andri Hardmeier}\footnote{Electronic address: andri@physik.unizh.ch}$^{1}$, 
 \mbox{Enrico Lunghi}\footnote{Electronic address: lunghi@physik.unizh.ch}$^{1}$, 
 \mbox{Dan Pirjol}\footnote{Electronic address: dpirjol@pha.jhu.edu}$^{2}$, 
 \mbox{and Daniel Wyler}\footnote{Electronic address: wyler@physik.unizh.ch}}

\affiliation{ 
Institut f\"ur Theoretische Physik, Universit\"at Z\"urich, \\ 8057 Z\"urich, Switzerland \\ 
$^{2}$ Department of Physics, The Johns Hopkins University, 3400 N. Charles Street, Baltimore MD 21218, U.S.A.} 
 
\vspace{2cm}

\begin{abstract}
  We discuss the most general form of the leading power suppressed
  collinear operators in the soft-collinear effective theory.  Such
  operators appear in the description of power corrections to
  exclusive heavy flavor decays into energetic light hadrons.
  Reparameterization invariance in the SCET provides powerful
  constraints on the Wilson coefficients of the subleading collinear
  operators.  We present explicit results for the matrix elements of
  these operators on pseudoscalar and vector mesons, which are
  expressed in terms of twist-2 and twist-3 light-cone wave functions.
  We consistently include the effects of three-particle light-cone
  distribution amplitudes and find that their impact could be of
  phenomenological relevance.
\end{abstract} 
 
\maketitle 
\newpage 
 

\section{Introduction} 
The soft-collinear effective theory (SCET)~\cite{BFL,BFPS,BS,BPS} has
been proposed as a systematic framework for the study of processes
involving energetic light quarks and gluons. Possible applications
include the decays of heavy hadrons into light particles in the
kinematical regions where the final products are very energetic, and
hard scattering processes involving light hadrons, such as deep
inelastic scattering and exclusive hadron form factors at large
momentum transfer. SCET provides a natural framework for establishing
a systematic expansion in $\Lambda/Q$ where $\Lambda$ is the QCD scale
and $Q$ is the typical large energy of the particles involved.  In
particular, it provides a convenient tool to establish factorization
theorems and study power corrections.

The observation that lies at the basis of SCET is that all the
kinematical singularities appearing in these processes are connected
to the exchange of collinear and soft particles ("long distance
modes").  Therefore, an explicit description of these processes in
terms of these degrees of freedom offers the usual advantages of an
effective theory approach. On the one hand, the power counting is
greatly simplified since it can be performed at the operator level;
on the other hand, standard renormalization group techniques can be
easily applied.

For definiteness we consider the decay of a heavy quark of mass $m$
which emits light quarks and gluons of energy $Q \simeq m$. The
possible external states are the soft spectator (whose momentum is of
order $\Lambda$) and the decay products (light mesons and leptons),
in addition to the heavy quark which cannot give rise to singularities and
can be considered as an external source. By the Coleman-Norton
theorem~\cite{CoNo}, the infra-red singularities of Feynman diagrams
describing these amplitudes correspond to the propagation of on-shell
particles. The idea underlying SCET is to identify all the possible
on-shell modes that can appear in the initial and final state (and
that are, hence, responsible for all the infra-red singularities) and
to write the effective theory of their interactions as an expansion in
$\Lambda/Q$.

In the following we use the standard light-cone decomposition of momenta
\beq 
p^\mu = \frac12 n^\mu \bn \cdot p + \frac12 \bn^\mu n \cdot p + p_\perp^\mu
\equiv (p_+,p_-,p_\perp)\,, 
\eeq 
where $n$ and $\bn$ are light-cone vectors satisfying $n^2= 
\bn^2 = 0$, $n\cdot \bn =2$. In any given process $n$ and $\bn$ are chosen
to be aligned to the final state collinear momenta. We also introduce
the dimensionless parameter $\lambda$ that
will serve as the expansion parameter of SCET. Its precise definition is
discussed below.

In the description of heavy meson decays, there are three relevant
kinematical configurations.

{\bf a) Soft quarks ($q_s, h_v\footnote{$h_v$ is the usual heavy quark
    degree of freedom that, after removing the fast oscillating
    components, is equivalent to a soft field.}$) and gluons ($A^\mu_s
  $)} with momenta $p_s \simeq \Lambda = Q (\lambda,\lambda, \lambda)$
(where we defined $\lambda = \Lambda/Q$).  The propagation of these
modes is described by the non-perturbative regime of QCD ($p_s^2
\simeq \Lambda^2$) and is therefore incalculable. Therefore, the
exchange of soft particles can only be parameterized and results in the
non-factorizable contributions to heavy-to-heavy and heavy-to-light
form factors and the $B$ meson wave function.

{\bf b) Collinear quarks ($\xi_n$) and gluons ($A^\mu_c $)} with
momenta $p_c \simeq Q (1, \lambda^2,\lambda)$.  These modes appear in
the description of the constituents of a fast-moving light meson.
They are also non-perturbative and their exchanges can only be
parameterized in terms of the light-cone wave functions of the final
state mesons.

{\bf c) Hard collinear quarks ($ \Xi_n $) and gluons ($ A^\mu_{hc}$)}
with momenta $p_{hc} \simeq Q (1,\lambda,\sqrt{\lambda})$. These modes
are necessary to describe inclusive jets ({\it e.g.} the $X_s$ system
near the end-point region of the photon spectrum in the inclusive
$B\to X_s \gamma$ decay~\cite{BFL}) and interactions of soft fields
with collinear particles ({\it e.g.} the $B$-meson soft spectator
after being struck by the energetic photon in $B \to \gamma e
\nu$~\cite{Li,DS,LPW,BHLN}). Hard collinear modes have virtuality of
order $p_{hc}^2\simeq \Lambda Q \gg \Lambda^2$, can be integrated out
perturbatively and result in so-called jet-functions.

Further intermediate 'hard' collinear modes with momenta 
$Q(1,\lambda, \lambda)$ occur at tree level (for instance in the coupling
of a soft and a collinear field). It is not necessary to introduce
explicitely such fields, because they are connected to hard collinears
via a reparameterization transformation. In Refs.~\cite{ChayKim,MMPS},
three different invariances under changes in the light cone
vectors $n$ and $\bn$ were introduced:
\bea 
{\rm type\; I:} & & n^\mu\to n^\mu+ \Delta_\perp^\mu,\,\, \bn^\mu \to \bn^\mu \\ 
{\rm type\; II:} & & \bn^\mu\to \bn^\mu+ \epsilon_\perp^\mu,\,\, n^\mu \to n^\mu \\ 
{\rm type\; III:} & & n^\mu\to n^\mu\alpha ,\,\, \bn^\mu \to \bn^\mu/\alpha 
\eea
where $\Delta_\perp^\mu\sim O(\lambda)$, ($\epsilon_\perp^\mu,  
\alpha) \sim O(1)$ and $\lambda$ refers to the expansion parameter 
relevant to the theory considered. In our case with two types of 
collinear fields, there are actually two expansion parameters, 
$\sqrt{\Lambda/Q}$ and $\Lambda/Q$ for the collinear and hard 
collinear fields, respectively.  Thus we must have two invariances, 
related to the corresponding fields. In particular, the hard collinear 
sector must satisfy a type-I reparameterization invariance with 
$\Delta_\perp^\mu\sim O(\sqrt\lambda)$. This invariance connects the 
$(1,\lambda,\sqrt\lambda)$ and $(1,\lambda,\lambda)$ modes. 

From a technical point of view, it is always possible to completely
integrate out the hard collinear modes because in all applications
they always appear as internal modes. This is obvious in processes
like $B\to \gamma e \nu$. In more complicated situations like heavy-light
semileptonic decays $B\to \pi e\nu$, a two step procedure has been 
proposed \cite{B2pi}.
In the first step an effective theory is formulated (SCET-I) containing 
only usoft and hard collinear modes, which is matched into a second step
onto the final effective theory (SCET-II) containing the collinear and
soft modes. [In the alternative treatment of Ref.~\cite{HN} only the
modes a) and b) are introduced and one matches directly from QCD onto
SCET-II.] On the other hand, in inclusive decays (like
$B\to X_s \gamma$) one can always write the decay width using the
optical theorem and again hard collinear modes can be viewed as
internal particles.

For the purpose of this paper, we take the point of view that problems
associated with the integration over the hard collinear modes at
subleading order have been cleared and proceed to the analysis of
power suppressed contributions.  Once the process is specified, the
integration of the hard collinears gives, order by order in
$\alpha_s$, all the relevant SCET operators (that will involve only
soft and collinear modes). The matrix elements of these 
operators between initial and final states fall in two groups: those 
factorizable in terms of "conventional" form factors and light-cone 
wave functions, and others that require the introduction of new
non-perturbative objects. In particular, the light-cone wave functions enter
through matrix elements of SCET operators (involving two quarks and an
infinite number of gluons) between the vacuum and a meson state.

In order to trust this perturbative computation, one also has to show
that higher loop contributions will not introduce new non-perturbative
structures at a given order in $\lambda$. This can be achieved by
writing the most general set of operators allowed by gauge and
reparameterization invariance and showing that their matrix elements
factorise.

In this paper we classify all the possible SCET collinear operators 
that appear at leading and subleading order in $\lambda$ and compute 
their matrix
elements in terms of the usual light cone distribution amplitudes of 
pseudoscalar~\cite{fily1,fily2} and vector~\cite{babr,babrhand,babr2} mesons. 
These operators are necessary for the
SCET analysis of any process involving energetic light mesons. In
particular, we find that for decays involving transverse polarized
vector mesons, only subleading operators contribute and that, in this
case, it is not possible to neglect the contribution of three-particle
distribution amplitudes.

Note, finally, that there are two equivalent formulations of SCET,
that are usually denoted as hybrid~\cite{BFL,BFPS} and coordinate
space~\cite{BCDF,HN}, respectively. In the former, strongly
oscillating collinear modes are removed from the theory by a partial
Fourier transformation; in the latter the slowly varying soft fields
are multipole expanded. These manipulations are necessary in order to
fully expand the Lagrangian in powers of $\lambda$. A failure in
achieving correctly the complete $\lambda$-expansion would result in a
theory whose infrared behaviour does not reproduce the one of full
QCD~\cite{TE}. We will work in the hybrid formalism and show how to
readily translate our results into coordinate-space.

We start by summarising briefly the main ingredients of the
hybrid formulation of SCET. In Sec.~\ref{bilinears} we present the
complete list of leading and subleading operators built only of collinear
fields. In Sec.~\ref{strategy} and \ref{matrixelements} we show how to 
extract the matrix elements of the various SCET operators and present our
results. In Sec.~\ref{RPI} we show that reparameterization
invariance (RPI) gives strong constraints on the Wilson coefficients
of the subleading collinear operators. 
In Sec.~\ref{ward} we present a sample application for the
subleading collinear operators constructed in this paper, discussing
final state photon emission in weak annihilation. 
We check our findings for the matrix elements of these operators
by verifying an exact Ward identity; this illustrates the numerical impact of the
three-particle contributions which are usually neglected in practical
computations. We collect our conventions in Appendix \ref{notation},
and in Appendix \ref{wf} we give the parameters of a minimal set of
light cone wave functions satisfying the equations of motion in QCD
used in the numerical evaluations in Sec.~\ref{ward}.

\section{Subleading collinear operators in SCET} 
\label{bilinears}

In this Section, we present a complete set of collinear quark
operators $\bar\xi_n \cdots \xi_n$ up to subleading order in
$\lambda$. Such operators are important in their own right, as
interpolating fields for meson states, and as building blocks for
more complicated operators in the effective theory. In constructing a
complete basis of operators we use constraints from the collinear 
gauge invariance, Dirac structure of the collinear quark fields 
$\xi_n$, and from reparameterization invariance of the effective 
theory. A similar procedure was recently used in Ref.~\cite{ps1} to 
construct the complete set of heavy-to-light currents in SCET.

The operators to be constructed in this Section contain the collinear
quark $\xi_{n,p}$ and gluon $A_{n,q}$ fields, together with the
collinear covariant derivative $iD_{c}^\mu = {\cal P}^\mu + gA_{n}^\mu$.
In order to be able to perform the power counting of the various
contributions at the operator level, it is convenient to assign a
$\lambda$ scaling to the fields requiring the kinetic terms to be of
order $O(1)$. In this way, one obtains $\xi_n \sim \lambda$ and 
$A^\mu_c \sim (\lambda^2, 1, \lambda)$.

In the hybrid formulation of SCET, one achieves a correct $\lambda$
expansion by extracting the large Fourier modes from each field:
\bea
\phi_c (x) = \sum_{\tilde p_c} e^{-i\tilde p_c x}\phi_{c,\tilde p_c} (x) 
\eea
where $\tilde p_c = Q (0,1,\lambda)$, are labels and the new field
$\phi_{c,\tilde p_c}$ is responsible for
fluctuations with momenta of order $Q(\lambda^2,\lambda^2,\lambda^2)$.

It is convenient to introduce a ``label''
operator ${\cal P}^\mu$~\cite{BS} which acts on the collinear
fields and picks up their large momentum: ${\cal P}^\mu \, \xi_{n,p} =
(\frac{\bar n \cdot p}{2} n^\mu + p_\perp^\mu)\, \xi_{n,p}$ and 
${\cal P}^\mu$, respectively.  When acting on a
product of several fields, these operators give the difference between
the total label carried by the fields minus the total label of the
complex conjugated fields.

We will also use a special notation which associates a
momentum label index to an arbitrary product of collinear
fields. Our convention is
\bea
\label{label}
\chi_{n,\omega} \equiv [W^\dagger \xi_n]_\omega = 
[\delta(\omega-\bn\mcdot {\cal P}) W^\dagger \xi_n]\,,\quad
[W^\dagger iD_{\perp c} W]_\omega = 
[\delta(\omega-\bn\mcdot {\cal P}) W^\dagger iD_{\perp c} W]
\eea
where $\delta(\omega-\bn\cdot {\cal P})$ acts only inside the
square brackets. 

The collinear
operators that we write in the following can include a nontrivial
flavor structure. When required, this will be denoted by a superscript 
showing the quark flavours. Note, finally, that for each operator we can 
have the singlet and octet colour structure. We will write explicitly 
only the former since the matrix elements of any octet operator between 
the vacuum and a meson state vanishes.

At leading order in $\lambda$ there are only three independent
collinear operators, which can be chosen as
\bea 
{\cal J}_V(\vec\omega) &=& 
\bar\chi_{n,\omega_1} \frac{\bnslash}{2} \chi_{n,\omega_2} \, , \label{JV} \\ 
{\cal J}_A(\vec\omega) &=& 
\bar\chi_{n,\omega_1} \frac{\bnslash}{2}\gamma_5 \chi_{n,\omega_2} \, , \label{JA} \\ 
{\cal J}^\alpha_T(\vec\omega) &=& 
\bar\chi_{n,\omega_1} \frac{\bnslash}{2}\gamma_\perp^\alpha 
\chi_{n,\omega_2}\, , \label{JT} 
\eea 
where $\gamma_\perp^\alpha \equiv \gamma^\alpha 
- n^\alpha \bnslash/2 - \bn^\alpha \nslash/2$ and $\vec \omega = 
(\omega_1,\omega_2)$.  Their transformation properties under charge 
conjugation are 
\bea 
\label{jvtC} 
{\cal J}^{(ud)}_{V,T}(\omega_1,\omega_2) &\to& 
-{\cal J}^{(du)}_{V,T}(-\omega_2,-\omega_1)\,, \\
\label{jaC}
{\cal J}^{(ud)}_A(\omega_1,\omega_2) &\to&
{\cal J}^{(du)}_A(-\omega_2,-\omega_1) \,.
\eea

At subleading order in $\lambda$, the number of allowed structures is 
much larger. It is convenient to choose a basis of collinear operators 
with simple transformation properties under charge conjugation. We 
choose the following four chiral-even collinear operators 
\bea\label{V1} 
{\cal V}^\alpha_1(\vec\omega) &=& 
\left[\bar\xi_n\frac{\bnslash}{2} 
(i\Dslash_{\perp c})^\dagger W_n\right]_{\omega_1} 
\frac{1}{\bn\mcdot {\cal P}^\dagger} \gamma^\alpha \chi_{n,\omega_2} + 
\bar\chi_{n,\omega_1} \gamma^\alpha \frac{1}{\bn\mcdot {\cal P}} 
\left[ W_n^\dagger 
i\Dslash_{\perp c}\frac{\bnslash}{2} \xi_n \right]_{\omega_2}\, ,\\ 
\label{V2} 
{\cal V}^\alpha_2(\vec\omega) &=& 
\left[\bar\xi_n\frac{\bnslash}{2} 
(iD_{\perp c}^\alpha)^\dagger W_n\right]_{\omega_1} 
\frac{1}{\bn\mcdot {\cal P}^\dagger}  \chi_{n,\omega_2} + 
\bar\chi_{n,\omega_1}  \frac{1}{\bn\mcdot {\cal P}} 
\left[ W_n^\dagger 
iD_{\perp c}^\alpha\frac{\bnslash}{2} \xi_n \right]_{\omega_2}\,,\\ 
\label{A1} 
{\cal A}^\alpha_1(\vec\omega) &=& 
\left[\bar\xi_n\frac{\bnslash}{2} 
(i\Dslash_{\perp c})^\dagger W_n\right]_{\omega_1} 
\frac{1}{\bn\mcdot {\cal P}^\dagger} \gamma^\alpha\gamma_5 \chi_{n,\omega_2} + 
\bar\chi_{n,\omega_1} \gamma^\alpha\gamma_5 \frac{1}{\bn\mcdot {\cal P}} 
\left[ W_n^\dagger 
i\Dslash_{\perp c}\frac{\bnslash}{2} \xi_n \right]_{\omega_2}\,,\\ 
\label{A2} 
{\cal A}^\alpha_2(\vec\omega) &=& 
\left[\bar\xi_n\frac{\bnslash}{2} 
(iD_{\perp c}^\alpha)^\dagger W_n\right]_{\omega_1} 
\frac{1}{\bn\mcdot {\cal P}^\dagger} \gamma_5 \chi_{n,\omega_2} - 
\bar\chi_{n,\omega_1} \gamma_5 \frac{1}{\bn\mcdot {\cal P}} 
\left[ W_n^\dagger 
iD_{\perp c}^\alpha\frac{\bnslash}{2} \xi_n \right]_{\omega_2}\,, 
\eea 
and the three chiral-odd operators 
\bea 
\label{S} 
{\cal S} (\vec\omega) &=& 
\left[\bar\xi_n\frac{\bnslash}{2} 
(i\Dslash_{\perp c})^\dagger W_n\right]_{\omega_1} 
\frac{1}{\bn\mcdot {\cal P}^\dagger} \chi_{n,\omega_2} + 
\bar\chi_{n,\omega_1}  \frac{1}{\bn\mcdot {\cal P}} 
\left[ W_n^\dagger i\Dslash_{\perp c} \frac{\bnslash}{2} 
\xi_n\right]_{\omega_2}\,,\\ 
\label{P} 
{\cal P} (\vec\omega) &=& 
\left[\bar\xi_n\frac{\bnslash}{2} 
(i\Dslash_{\perp c})^\dagger W_n\right]_{\omega_1} 
\frac{\gamma_5}{\bn\mcdot {\cal P}^\dagger}  \chi_{n,\omega_2} + 
\bar\chi_{n,\omega_1}  \frac{\gamma_5}{\bn\mcdot {\cal P}} 
\left[ W_n^\dagger i\Dslash_{\perp c} \frac{\bnslash}{2} 
\xi_n\right]_{\omega_2} \,,\\ 
\label{T} 
{\cal T}^{\alpha\beta}(\vec\omega) &=& 
\left[\bar\xi_n\frac{\bnslash}{2} 
(iD_{\perp c}^\alpha)^\dagger W_n\right]_{\omega_1} 
\hskip -0.16 cm 
\frac{\gamma_\perp^\beta}{\bn\mcdot {\cal P}^\dagger}  \chi_{n,\omega_2} - 
\bar\chi_{n,\omega_1}  \frac{\gamma_\perp^\beta}{\bn\mcdot {\cal P}} 
\left[ W_n^\dagger iD_{\perp c}^\alpha \frac{\bnslash}{2} 
\xi_n\right]_{\omega_2} \,, 
\eea 
together with the corresponding colour octet operators which will be 
denoted by the same letter and a colour index $a$. We include the 
factors $1/\bn\mcdot {\cal P}$ and $1/\bn\mcdot {\cal P}^\dagger$ in 
the definition of the operators, Eqs.~(\ref{V1})--(\ref{T}), to make 
them invariant under the transformation $n\to n\alpha$, $\bn\to \bn / 
\alpha$ (type-III reparameterization invariance). 
 
These operators are not the most general collinear gauge invariants
at $O(\lambda)$. In analogy to the heavy-to-light current 
considered in Ref.~\cite{ps1}, it is possible to write also 
three-particle operators, which contain three collinear gauge 
invariant factors.  Their Dirac structure is again restricted by the 
effective theory constraint $\nslash\xi_n = 0$, which leaves two 
possible chiral-even operators 
\bea 
{\cal V}_3^\alpha(\vec\omega) &=& \bar \chi_{n, \omega_1} \, 
\frac{\bnslash}{2}\,  [\frac{1}{\bn\mcdot {\cal P}}W^\dagger iD_\perp^\alpha W ]_{\omega_3} \, 
\chi_{n,\omega_2} =
\bar\chi_{n,\omega_1} \frac{\bnslash}{2}\left[ \left(\frac{1}{\bn\mcdot {\cal P}}\right)^2
W^\dagger ig\bar n_\beta G^{\beta\alpha} W\right] \chi_{n,\omega_2} \\
{\cal A}_3^\alpha(\vec\omega) &=& \bar \chi_{n, \omega_1}  \,
\frac{\bnslash}{2}\gamma_5 \, 
[\frac{1}{\bn\mcdot {\cal P}} W^\dagger iD_\perp^\alpha W ]_{\omega_3} \, 
\chi_{n,\omega_2} =
\bar\chi_{n,\omega_1} \frac{\bnslash}{2}\gamma_5
\left[ \left(\frac{1}{\bn\mcdot {\cal P}}\right)^2
W^\dagger ig\bar n_\beta G^{\beta\alpha} W\right] \chi_{n,\omega_2} \nn\\
\eea 
and a single chiral--odd operator 
\bea 
{\cal T}_3^{\alpha\beta} (\vec\omega) &=&   \bar \chi_{n, \omega_1} \, 
\frac{\bnslash}{2} \, \gamma^\alpha_\perp\,  [\frac{1}{\bn\mcdot {\cal P}} 
W^\dagger iD_\perp^\beta W ]_{\omega_3}  \,
\chi_{n,\omega_2} =
\bar\chi_{n,\omega_1} \frac{\bnslash}{2} \gamma^\alpha_\perp
\left[ \left(\frac{1}{\bn\mcdot {\cal P}}\right)^2
W^\dagger ig\bar n_\rho G^{\rho\beta} W\right] \chi_{n,\omega_2}
 \, . \nn\\
\label{t3ab} 
\eea 
The factor $1/\bn\mcdot {\cal P}$ assures again invariance under 
type-III reparameterization invariance. 
 
The transformation properties of the subleading operators under charge 
conjugation are 
\bea\label{cc1} 
{\cal V}^{(ud)\alpha}_{1,2}(\omega_1,\omega_2) &\to & 
-{\cal V}^{(du)\alpha}_{1,2}(-\omega_2,-\omega_1)\,, \\ 
{\cal A}^{(ud)\alpha}_{1,2}(\omega_1,\omega_2) & \to & 
{\cal A}^{(du)\alpha}_{1,2}(-\omega_2,-\omega_1)\,, \\ 
{\cal S}^{(ud)\alpha}(\omega_1,\omega_2) &\to & 
{\cal S}^{(du)\alpha}(-\omega_2,-\omega_1) \,,\\ 
{\cal P}^{(ud)\alpha}(\omega_1,\omega_2) &\to& 
-{\cal P}^{(du)\alpha}(-\omega_2,-\omega_1)\,, \\ 
{\cal T}^{(ud)\alpha\beta}(\omega_1,\omega_2) &\to& 
 {\cal T}^{(du)\alpha\beta}(-\omega_2,-\omega_1)\,. 
\label{cc2} 
\eea 

The corresponding operators of opposite charge conjugation properties 
can be constructed by changing the relative sign of the two terms in 
Eqs.~(\ref{V1})-(\ref{T}). They will be denoted with a tilde, e.g. 
\bea\label{V1tilde} 
\widetilde{\cal V}^\alpha_1(\vec\omega) &=& 
\left[\bar\xi_n\frac{\bnslash}{2} 
(i\Dslash_{\perp c})^\dagger W_n\right]_{\omega_1} 
\frac{1}{\bn\mcdot {\cal P}^\dagger} \gamma^\alpha \chi_{n,\omega_2} - 
\bar\chi_{n,\omega_1}\gamma^\alpha \frac{1}{\bn\mcdot {\cal P}} 
\left[ W_n^\dagger 
i\Dslash_{\perp c}\frac{\bnslash}{2} 
\xi_n\right]_{\omega_2}\,, 
\eea 
and similarly for the remaining 6 operators. They are not independent 
and can be related to the above ones by using the Dirac identities in the 
Appendix A. For example, using Eq.~(\ref{id2}) (written in terms of the
transverse antisymmetric tensor $\varepsilon^\perp_{\mu\nu} \equiv \frac12
\varepsilon_{\mu\nu\alpha\beta} \bn^\alpha n^\beta$)
\bea 
\frac{\bnslash}{2}\gamma^\mu_\perp \gamma^\nu_\perp = 
g_\perp^{\mu\nu} \frac{\bnslash}{2} - i\varepsilon_\perp^{\mu\nu} 
\frac{\bnslash}{2}\gamma_5 \,, 
\eea 
one obtains the relations 
\bea\label{dirac} 
{\cal V}_1^\mu &=& {\cal V}_2^\mu + i\varepsilon_\perp^{\mu\nu} 
\widetilde{\cal A}_{2\nu}\,,\qquad 
{\cal A}_1^\mu = {\cal A}_2^\mu +  i\varepsilon_\perp^{\mu\nu} 
\widetilde{\cal V}_{2\nu}  \, ,     \\ 
\widetilde {\cal V}_1^\mu &=& \widetilde {\cal V}_2^\mu + 
 i\varepsilon_\perp^{\mu\nu} 
{\cal A}_{2\nu}\,,\qquad 
\widetilde {\cal A}_1^\mu = \widetilde {\cal A}_2^\mu + 
 i\varepsilon_\perp^{\mu\nu} {\cal V}_{2\nu} \nn \, .
\eea 
They can be solved for $\widetilde {\cal V}_i$, 
$\widetilde {\cal A}_i$, with the results
\bea\label{VAtilde} 
& & \widetilde {\cal V}_1^\alpha = i\varepsilon_\perp^{\alpha\beta}{\cal A}_{1\beta}\,, 
\hskip 2.35cm 
\widetilde {\cal A}_1^\alpha = i\varepsilon_\perp^{\alpha\beta}{\cal V}_{1\beta}\,, 
\label{tilde}   \\ 
& & \widetilde {\cal V}_2^\alpha = i\varepsilon_\perp^{\alpha\beta} 
({\cal A}_{1\beta} - {\cal A}_{2\beta})    \, , 
 \qquad 
\widetilde {\cal A}_2^\alpha = i\varepsilon_\perp^{\alpha\beta} 
({\cal V}_{1\beta} - {\cal V}_{2\beta})\,. \label{ww} 
\eea 

For the chiral-odd operators there are two identities following
from their definition
\bea\label{StildeT}
g^\perp_{\alpha\beta} {\cal T}^{\alpha\beta}(\vec\omega) = \tilde S(\vec\omega)\,,
\qquad
g^\perp_{\alpha\beta} \tilde {\cal T}^{\alpha\beta}(\vec\omega) = S(\vec\omega)\,,
\eea
and two other identities following from Eq.~(\ref{id4})
\bea\label{PtildeT}
{\cal P} = i\varepsilon_{\mu\nu}^\perp \tilde {\cal T}^{\mu\nu}\,,\qquad
\tilde {\cal P} = i\varepsilon_{\mu\nu}^\perp {\cal T}^{\mu\nu}\,.
\eea

Let us finally discuss the way these operators appear in explicit
calculations.  A given QCD operator ${\cal O}_{\rm QCD}$ is matched
onto SCET operators containing the subleading collinear bilinears
introduced above
\bea\label{OQCD}
{\cal O}_{\rm QCD} &=& \cdots + \int {\rm d}\omega_1 {\rm d}\omega_2 C_1(\omega_1,\omega_2)
\{\cdots \} {\cal V}_i(\omega_1,\omega_2)\\
&+& \int {\rm d}\omega_1 {\rm d}\omega_2 {\rm d}\omega_3 C_2(\omega_1,\omega_2,\omega_3)
\{\cdots \} {\cal V}_i(\omega_1,\omega_2,\omega_3)\nn
\eea
where the ellipses $\{\cdots \}$ denote possible soft fields which
were omitted in writing the SCET operator. The Wilson coefficients
$C_{1,2}(\omega_i)$ depend on the momentum labels of the collinear
bilinears.

After factorization, the matrix elements of the collinear operators 
${\cal V}_i(\omega_i)$ between a light meson and vacuum
lead to non-perturbative functions $\langle M(p_M) |{\cal V}_i(\omega_1,\omega_2)|0
\rangle \simeq \varphi_i(u)$. It is convenient to implement momentum
conservation $\omega_1 - \omega_2 = p_M$ by introducing the momentum fraction
$u$ by $(\omega_1,\omega_2) = (u,-\bar u) \bn\cdot p_M$, with $u=(0,1)$.
The charge-conjugation transformation properties of the
collinear operators ${\cal V}_i(\omega_1,\omega_2)$ (see
Eqs.~(\ref{jvtC}),(\ref{jaC}),(\ref{cc1})-(\ref{cc2})), taken together
with the $C$ quantum number of the state $|M(p_M)\rangle$, fixes
the symmetry property of matrix elements under the substitution $u\to
\bar u$.  For example, taking $C=-1$ as appropriate for the $\rho$
meson, one has
\bea
\langle \rho(p,\eta)|{\cal V}_i^{\rm even (odd)}(\omega_1,\omega_2)|0\rangle \sim
\varphi^{\rm odd (even)}_\rho(u)\,,
\eea
such that only the odd (even) part of the corresponding Wilson
coefficient $C(\omega_1,\omega_2)$ will give a non-vanishing
contribution to the given matrix element of Eq.~(\ref{OQCD}).
The matrix elements of Eq.~(\ref{OQCD}) will be given with the 
integration measure $\Pi_i \mbox{d}\omega_i$ replaced with $\mbox{d}u$ for the 
2-parton operators, and with $\mbox{d}\alpha_1 \mbox{d}\alpha_2 \mbox{d}\alpha_3 
\delta(1-\Sigma_{i=1}^3 \alpha_i)$ for the 3-parton operators.

\section{Strategy for the computation of the matrix elements} 
\label{strategy}

In this section we explain the procedure that we use to extract the 
matrix elements between the vacuum and a state with one vector or 
pseudo--scalar meson of the leading and subleading collinear operators 
introduced in Sec.~\ref{bilinears}. We will see that they can be 
expressed in terms of the vector and pseudo--scalar mesons' light-cone 
wave functions. 
 
The usual starting point are matrix elements of nonlocal operators of 
the form
\beq
\langle M | \, \bar q(x)\, \Gamma \, {\cal W}[x,y] \, q(y) \, | 0 \rangle 
\quad {\rm or} \quad 
\langle M | \, \bar q(x)\, \Gamma \, {\cal W}[x,z]\,  G_{\mu\nu} (z)\, 
{\cal W}[z,y] \, q(y)\, | 0 \rangle \,. 
\eeq 
The ${\cal W}[\cdot , \cdot]$ terms are Wilson lines required by gauge 
invariance and are explicitly given by 
\beq 
{\cal W}[x,y] = P\exp 
\left[ i g \, (x-y)^\mu \int_0^1 {\rm d }t A^\mu (t x + \bar{t} y) 
\right] \,, 
\label{qcdW} 
\eeq 
where $\bar{t} = 1-t$. In the following, the presence of the 
appropriate Wilson lines between fields evaluated at different 
space-time points is understood. 
 
The general procedure that we adopt is rather simple. We first project 
the QCD operators onto SCET keeping leading and 
subleading contributions; then we use the standard definitions of the 
light-cone wave functions given in Refs.~\cite{fily1,fily2,babr,babr2,BeFe} 
to identify the matrix elements of the various SCET operators. 
 
The presence of Wilson lines and of fields situated at different 
space-time positions, causes the appearance of arbitrarily suppressed 
operators in the projection of the QCD operators onto SCET. In order 
to simplify the extraction of the leading and subleading operators we 
expand the QCD operators around the transverse direction: 
\bea 
\bar q(x) \, \Gamma \, q(y) & = & 
\Big. \bar q(x) \,  \Gamma \, q(y)\Big|_{ 
{ }^{x_\perp = 0}_{y_\perp = 0}} 
+ x_\perp^\mu \, \Big. {\partial \over \partial x_\perp^\mu} 
\bar q(x)  \, \Gamma \, q(y) \Big|_{ 
{ }^{x_\perp = 0}_{y_\perp = 0}} \nn \\ 
& & + y_\perp^\mu \, \Big. {\partial \over \partial y_\perp^\mu} 
\bar q(x)  \, \Gamma \, q(y)\Big|_{ 
{ }^{x_\perp = 0}_{y_\perp = 0}} 
+ \hbox{higher powers of $x_\perp$, $y_\perp$} \, ,\label{exp} 
\eea 
where 
\bea 
{\partial \over \partial x_\perp^\mu} \bar q(x)  \, \Gamma \, q(y) 
& = & 
\bar q (x) \overleftarrow{D}_{\perp\mu}  \, \Gamma \, q(y) - 
i \int_0^1 {\rm dt} \, t \, z^\alpha \, 
\bar q (x) g G_{\alpha \mu} (t x  +  \bar t y) \, \Gamma \, q(y) \, , 
\\ 
{\partial \over \partial y_\perp^\mu} \bar q(x)  \, \Gamma \, q(y) 
& = & 
\bar q (x)  \, \Gamma \, D_{\perp\mu} q(y) - 
i \int_0^1 {\rm dt} \, \bar t \, z^\alpha \, 
\bar q (x) g G_{\alpha \mu} ( t x +  \bar t y) \, \Gamma \, q(y) \, , 
\eea 
and we define $z = x-y$. 
 
All the terms of this expansion count as $O(1)$ (in fact, we only 
require $x-y$ to be on the light-cone without any constraint on the 
direction of this vector) but the projections of the various terms 
start with SCET operators of increasing $\lambda$ suppression: 
\bea 
\Big. \bar q(x) \, {\cal W}[x,y] \, \Gamma \, q(y)\Big|_{ 
{ }^{x_\perp = 0}_{y_\perp = 0}} 
&\to& {\cal O}_0 + {\cal O}_1 + \cdots \, , \nn  \\ 
\Big. {\partial \over \partial x_\perp^\mu} 
\bar q(x) \, {\cal W}[x,y] \, \Gamma \, q(y) \Big|_{ 
{ }^{x_\perp = 0}_{y_\perp = 0}} 
&\to& {\cal O}_1 + {\cal O}_2  + \cdots \, ,  \\ 
\Big. {\partial \over \partial x_\perp^\mu \partial x_\perp^\nu} 
\bar q(x) \, {\cal W}[x,y] \, \Gamma \, q(y) \Big|_{ 
{ }^{x_\perp = 0}_{y_\perp = 0}} 
&\to& {\cal O}_2 + {\cal O}_3  + \cdots \, , \nn 
\eea 
where ${\cal O}_n$ are generic SCET operators suppressed by a factor 
$\lambda^n$ compared to ${\cal O}_0$. Since we are interested in 
leading and subleading operators only, we just need to consider the 
few terms explicitly written in Eq.~(\ref{exp}). 
 
Let us now show in detail how to perform the SCET projection of a 
given QCD operator. The starting point is to express the QCD quark 
field in terms of collinear and soft fields 
\bea 
q(x) &\to& \sum_{\tilde p_c} e^{- i \tilde p_c \cdot x} 
\left[ 1 + {1\over i \bn \cdot D_c} i \Dslash_{\perp c} 
{\bnslash \over 2} \right] \xi_{n,\tilde p_c} (x) + 
\sum_{\tilde p_s}  e^{- i \tilde p_s \cdot x} q_{\tilde p_s} (x) \, .
\eea 
Since all the terms in the expansion (\ref{exp}) are evaluated at 
$x_\perp = 0$ we obtain
\bea 
\label{replacement} 
q(x) &\to& \sum_{\bn \cdot \tilde p_c}
e^{- {i\over 2} \bn \cdot \tilde p_c n \cdot x} 
\left(1 + {1\over i \bn \cdot D_c} i \Dslash_{\perp c} 
{\bnslash \over 2} \right) \xi_{n,\tilde p_c} (x)  \, ,
\eea 
where we do not include the soft quark because, as argued below, its
contribution receives an additional power suppression and it is
negligible for the analysis of leading and subleading operators.

Let us now consider the leading order projection of the QCD operator
$\bar q(x) \, {\cal W}[x,y]\, \gamma^\mu \, q(y)$. The matrix element
of the $\bar\xi\xi$ term is given by
\bea 
\label{1st} 
\langle M(p_M) | 
\int {\cal D}^2\vec \omega \, 
\bar \chi_{n,\omega_1} \, n^\mu{\bnslash\over 2 }\,  \chi_{n, \omega_2} |0\rangle 
&=& \, \int_0^1 {\rm d}u \, 
e^{i (u \, p_M \cdot x + \bar u p_M \cdot y)} \, 
 \langle M | \bar \chi_{n, u p_M} \, n^\mu{\bnslash\over 2 }\, \chi_{n, -\bar u p_M} |0\rangle  \quad 
\\ 
& \propto &  n_\mu \int_0^1 {\rm d}u \, 
e^{i (u \, p_M \cdot x + \bar u p_M \cdot y)} \, 
 \varphi (u) \, ,\nn
\eea 
where 
\beq 
\int {\cal D}^2 \vec{\omega} \equiv \int 
{\rm d}\omega_1 \, {\rm d}\omega_2  \, 
e^{{i\over 2} (\omega_1 \, n\cdot x - \omega_2 \, n\cdot y)}  \,, 
\label{d2omega} 
\eeq 
and $\bar u = 1-u$. The $W_n$ factors (contained in the $\chi_n$ fields)
come from the projection of the full QCD Wilson line, Eq.(\ref{qcdW}),
in the limit $x_\perp=0=y_\perp$.  In Eq.~(\ref{1st}), we replaced the
discrete sum of Eq.~(\ref{replacement}) by the integral over
$\omega_i$.  Once momentum conservation is imposed, the sum of the two
labels must be equal to the total momentum of the meson:
$\omega_1-\omega_2 = \bn \cdot p_M$ and the fractions of momentum
carried by the two collinear quarks are $\omega_1= u p_M$ and
$\omega_2=- \bar u p_M$, respectively (this assumes that the parton
carrying momentum $\omega_1 (\omega_2)$ ends up as a quark (antiquark) in $M$). 
The equality in Eq.~(\ref{1st}) is obtained by direct comparison with the 
definition of the QCD light-cone wave functions for the $\bar q \gamma^\mu q$
current. The result is
\bea 
\langle M | \bar \chi_{n, \omega_1}  \, n^\mu{\bnslash\over 2 }\, \chi_{n, \omega_2} |0\rangle 
& \propto & \varphi (u) \, . 
\eea 
Note that although in principle mixed soft-collinear terms $\bar
q\xi_n$ can be generated too, their matrix elements on a collinear
meson state require insertions of the soft-collinear subleading
Lagrangian and are suppressed relative
to those of the diagonal operators $\bar\xi_n \xi_n$ we consider here.

Before concluding this section, let us explain how to translate the 
above arguments into the coordinate-space formulation of SCET. In the 
first place note that our leading and subleading operators do not 
involve soft covariant derivatives. This implies that each operator 
can be trivially translated into coordinate-space formalism by 
replacing the label operators with ordinary derivatives: ${\cal 
  P}^\mu_\perp + g\, A^\mu_{c\perp} \to i \partial_\perp^\mu + g\, 
A^\mu_{c\perp}$ and $\bn \cdot {\cal P} \to i \bn \cdot \partial$. The 
latter substitution might rise concerns about the invariance of the 
coordinate-space version of Eqs.~(\ref{V1})--(\ref{t3ab}) under soft 
gauge transformations~\cite{BCDF,BF}. From the discussion in 
Ref.~\cite{BF} it follows that the expansion in $\lambda$ is well 
defined only if the gauge transformations of the fields are 
homogeneous ({\it i.e.} the theory must be gauge invariant order by 
order in $\lambda$). This can always be achieved via appropriate 
redefinitions of the collinear fields. In Ref.~\cite{BF} it has been 
shown that homogeneous soft gauge transformations cannot depend on 
$\bn\cdot x$, hence, $i \bn \cdot \partial$ derivatives can be 
inserted without spoiling soft gauge invariance. From these 
considerations it becomes clear that matrix elements of the 
coordinate-space version of our operators will have the form
\bea 
\langle M(p_M)  | \bar \xi_{n}(x)  \Gamma' \xi_{n}(y) |0\rangle 
& \propto & \int_0^1 {\rm d}u \, 
e^{i (u \, p_M \cdot x + \bar u p_M \cdot y)} \,\varphi (u)\, . 
\eea 

\section{Computation of the matrix elements} 
\label{matrixelements} 

In this section we apply the technique introduced above to extract the 
matrix elements of the leading and subleading SCET operators listed in 
Sec.~\ref{bilinears}. First we use Eq.~(\ref{exp}) to project the 
three- and two-particle QCD operators onto SCET.  Then we compare the 
resulting expansion with the definition of the light-cone wave 
functions of pseudo-scalar and vector mesons and extract the SCET 
matrix elements. 

\subsection{SCET decomposition of the QCD operators} 
Let us now consider the SCET decomposition of the five independent QCD 
currents and of the relevant three-particle operators, respectively. 
The matrix elements of the vector, axial and tensor currents involve 
twist-2 and twist-3 light-cone wave functions and it is therefore 
necessary to consider the full expansion in Eq.~(\ref{exp}). On the 
other hand, the matrix elements of the scalar, pseudo-scalar and 
three-particle operators involve only twist-3 and higher wave 
functions and we need to keep only the leading term in 
Eq.~(\ref{exp}). 
 
The SCET decomposition of the various currents are (with
$z=x-y$) 
\bea \label{scs} 
\hskip 0cm 
\bar{q}(x)  q(y) &= & 
  \int {\cal D}^2 \vec\omega \, 
        {\cal S}    (\vec \omega )  \, ,   \hskip 11.5cm 
\eea 
\bea\label{scp}
    \bar{q}(x)  \gamma^5 q(y)& =& 
      \int {\cal D}^2 \vec\omega \, 
            {\cal P}    (\vec \omega )   \, ,       \hskip 11cm 
\eea 
\bea\label{scv}
       \bar{q}(x)\gamma^{\mu}  q(y)&=& 
   \int {\cal D}^2 \vec\omega \, 
    \bigg\{  {\cal J}_{V}    (\vec \omega ) n^{\mu} 
   + {\cal V}_{1}^{\mu}(\vec \omega )    \hskip 8.6cm  \nn \\ 
  & & \hskip 0cm + {i\over 2} \, n^{\mu} \left[ (\omega_1 \, x_{\perp\alpha} 
   -   \omega_2 \,  y_{\perp\alpha}) \, {\cal V}_{2}^{\alpha}(\vec \omega ) 
  + (\omega_1 \, x_{\perp\alpha} 
   +   \omega_2 \,  y_{\perp\alpha})  \, 
  \widetilde{\cal V}_{2}^{\alpha}(\vec \omega ) \right] \bigg\}  \nn \\ 
  & & - {1\over 2} \, n^{\mu}  n \cdot z 
   \int_{0}^{1} {\rm d}t \, (t x_{\perp\alpha} +  \bar{t} y_{\perp\alpha}) 
   \int \cD^3 \vec{\omega} \,\omega_{3}^{2} \, {\cal V}_{3}^{\alpha}(\vec \omega )\, , 
\eea 
\bea\label{sca}
\bar{q}(x)\gamma^{\mu} \gamma^5 q(y)&=& 
   \int {\cal D}^2 \vec\omega \, 
    \bigg\{  {\cal J}_{A}    (\vec \omega ) n^{\mu} 
   + {\cal A}_{1}^{\mu}(\vec \omega )   \hskip 8.3cm \nn \\ 
  & & \hskip 0cm + {i\over 2} \, n^{\mu} \left[ (\omega_1 \, x_{\perp\alpha} 
   -   \omega_2 \,  y_{\perp\alpha}) \, {\cal A}_{2}^{\alpha}(\vec \omega ) 
  + (\omega_1 \, x_{\perp\alpha} 
   +   \omega_2 \,  y_{\perp\alpha})  \, 
  \widetilde{\cal A}_{2}^{\alpha}(\vec \omega ) \right] \bigg\}  \nn \\ 
  & & - {1\over 2} \, n^{\mu}  n \cdot z 
   \int_{0}^{1} {\rm d}t \, (t x_{\perp\alpha} +  \bar{t} y_{\perp\alpha}) 
   \int \cD^3 \vec{\omega} \,\omega_{3}^{2} \, {\cal A}_{3}^{\alpha}(\vec \omega )\, , 
\eea 
\bea\label{sct}
\bar{q}(x)\sigma_{\mu \nu} q(y)&=& 
\int {\cal D}^2  \vec{\omega}\Big\{   i \left(n_{\mu} g_{\perp\alpha \nu}-n_{\nu}g_{\perp\alpha \mu}\right) 
{\cal J}_{T}^{\alpha}(\vec{\omega}) 
+ {i\over 2} \left(\bar{n}_{\mu}\,n_{\nu}-n_{\mu}\, \bar{n}_{\nu} \right)
\widetilde{{\cal S}}(\vec{\omega}) 
+  i {\cal T}_{[\mu \nu]}(\vec{\omega})\nn \\ 
& &  \hskip -1cm 
- {1\over 2} 
(n_{\mu} g_{\perp\nu\beta}-n_{\nu}g_{\perp\mu \beta} ) \left[ 
(\omega_1 \, x_{\perp\alpha} - \omega_2 \, y_{\perp\alpha}) \, {\cal T}^{\alpha \beta}(\vec \omega ) 
+ (\omega_1 \, x_{\perp\alpha}+ \omega_2 \, 
y_{\perp\alpha}) \widetilde{\cal T}^{\alpha \beta}(\vec \omega )\right] \Big\} \nn \\ 
& & -i \,  n \cdot z 
( n_\mu g_{\perp\nu \gamma} - n_\nu g_{\perp\mu \gamma} ) 
\int_0^1 {\rm d}t \, (t x_{\perp\alpha} +  \bar{t} y_{\perp\alpha}) 
\int \cD^3 \vec{\omega} \, {\omega_{3}^{2}\over 2} \, 
{\cal T}_{3}^{\gamma\alpha}(\vec \omega )  \, , 
\label{te}\eea 
where $g_{\perp}^{\mu \nu} \equiv 
g^{\mu\nu}-\frac12 n^{\mu}\bar{n}^{\nu}-\frac12 n^{\nu}  \bar{n}^{\mu}$, and
\beq
\int {\cal D}^3 \vec{\omega} \equiv \int
 {\rm d}\omega_1\,  {\rm d}\omega_2 \,{\rm d}\omega_3 \, 
e^{{i\over 2} (\omega_1 \, n\cdot x + \omega_3 n\cdot (tx+\bar t y)- \omega_2 \, n\cdot y)}\, . 
\label{d3omega} 
\eeq 

There are only three 3-particle operators whose matrix elements 
involve twist-3 light-cone wave functions~\cite{babr,babr2}: 
\bea 
\bar q (x) \, g G_{\mu\nu}(tx+\bar t y) \, \gamma_\alpha \, q(y) & = & 
-i \, n_\alpha \left[ n_\mu g_{\perp\nu \gamma} - n_\nu g_{\perp\mu \gamma} \right] 
\int {\cal D}^3 \vec{\omega} \, {\omega_3^2 \over 2} \, {\cal V}_3^\gamma (\vec{\omega}) \,,\\ 
\hskip -1.6cm 
\bar q (x) \, g G_{\mu\nu}(tx+\bar t y) \, \gamma_\alpha  \gamma_5 \, q(y) & = & 
-i \, n_\alpha \left[ n_\mu g_{\perp\nu \gamma} - n_\nu g_{\perp\mu \gamma} \right] 
\int {\cal D}^3 \vec{\omega} \, {\omega_3^2 \over 2} \, {\cal A}_3^\gamma (\vec{\omega}) \,,\\ 
\bar q (x) \, g G_{\mu\nu}(tx+\bar t y) \, \sigma_{\alpha \beta} \, q(y) & = & 
\left[ n_\mu g_{\perp\nu \gamma} - n_\nu g_{\perp\mu \gamma} \right] 
           [ n_\alpha g_{\perp\beta \sigma} - n_\beta g_{\perp\alpha \sigma} ] 
\int  {\cal D}^3 \vec{\omega} \, {\omega_3^2 \over 2} \, 
     {\cal T}_3^{\sigma\gamma} (\vec{\omega}) \,.\;\;\;\;\; \label{3te} 
\eea

\subsection{Pseudo-scalar meson matrix elements}

In the following we collect the matrix elements between the vacuum and
a pseudo-scalar meson state of all the relevant QCD operators \cite{fily1,fily2},
including the contributions from all wave functions of twist 2 and 3:
\bea
\langle P(p)|\bar q(x)\, q(y)|0\rangle &=& 0  \,,  \\
\langle P(p)|\bar q(x)\gamma^5 q(y)|0\rangle &=&
-i\,f_{P}\mu_{P}\int_0^1 {\rm d}u\,e^{i(u\,p\cdot x+\bar{u}\,p\cdot y)}
\varphi_{p}(u)  \,, \\
\langle P(p)|\bar q(x)\gamma^{\mu} q(y)|0\rangle &=& 0 \\
\langle P(p)|\bar q(x)\gamma^{\mu} \gamma^5 q(y)|0\rangle &=&
 -i f_{P} p^{\mu}
\int_0^1 {\rm d}u\,e^{i(u\,p\cdot x+\bar{u}\,p\cdot y)} \varphi(u)\,,  \\
\langle P(p)|\bar q(x)\sigma^{\mu \nu}\gamma_5 q(y)|0\rangle &=&
i \frac{f_{P}\mu_P}{6} (p^\mu z^\nu - p^\nu z^\mu) 
\int_0^1 {\rm d}u\,e^{ i(u\,p\cdot x+\bar{u}\,p\cdot y)} \varphi_{\sigma}(u) \\
& &  \hskip -5cm
\langle P(p)|\bar q(x)\sigma^{\mu\nu}\gamma_5\, 
g\,G^{\alpha \beta}(t x+\bar t y) q(y)|0\rangle\\
&=&
i[ p_\mu (p_\alpha g_{\nu\beta} - p_\beta g_{\alpha \nu} - (\mu\leftrightarrow \nu)]
f_{3\pi}
\int {\cal D}^3\vec{\alpha} \varphi_{3 \pi}(\vec{\alpha})\nn
\eea
where $f_P$ is the decay constant, $\mu_P = m_P^2 /(m_1+m_2)$ ($m_i$ 
denote the masses of the light valence quarks), $f_{3\pi}$ is the
three-particle decay constant (with dimension GeV$^2$). 
The 3-body integration measure is defined as
\bea\label{d3alpha}
\int {\cal D}^3\vec{\alpha} \equiv \int_0^1 {\rm d} \alpha_1\, 
{\rm d} \alpha_2 \, {\rm d} \alpha_3  \, 
e^{i (\alpha_1 p\cdot x + \alpha_3 p\cdot ( t x+ \bar t y) + \alpha_2 p\cdot y)} 
\delta (1-\sum \alpha_i) \, .
\eea
The wave function $\varphi (u)$ is of twist-2 while $\varphi_p (u)$, 
$\varphi_\sigma (u)$ and $\varphi_{3\pi} (\vec\alpha)$ are twist-3. 
In the matrix element of the tensor current one can use partial
integration to remove an explicit $n\cdot z$ factor, which gives
\bea
\langle P(p)|\bar q(x)\sigma^{\mu \nu} q(y)|0\rangle
&=& { f_{P}\mu_{P}\over 6} \int_0^1 {\rm d}u\,e^{ i(u\,p\cdot x+\bar{u}\,p\cdot y)}
\Big[
{\bar{n} \cdot p\over 2} \,\epsilon^{\mu \nu \rho \tau}n_{\rho}z_{\tau\perp}  
\varphi_{\sigma}(u)
- i \epsilon^{\mu \nu}_{\perp}{\varphi_{\sigma}}^{\prime}(u)
\Big]   \,, \;\;\;\;   
\eea
Also, the matrix element of the 3-parton operator can be written as
\bea
\langle P(p)|\bar q(x)\sigma^{\alpha \beta}\, g\,G^{\mu \nu}
(t x+\bar t y) q(y)|0\rangle =
{f_{3 \pi} \over 4} (\bar{n} \cdot p)^2  n_{\rho}
\left(n^{\mu}\, \varepsilon^{\alpha \beta \rho \nu}-n^{\nu}\, 
\varepsilon^{\alpha \beta \rho \mu} \right)
\int \hskip -0cm {\cal D}^3\vec{\alpha} \varphi_{3 \pi}(\vec{\alpha})\,,
\eea

From the
comparison with Eqs.~(\ref{scs})-(\ref{3te}) we can easily extract the
matrix elements of all the SCET operators introduced in Sec.~\ref{bilinears}:
\bea\label{scet1p}
\langle P(p) |{\cal J}_A (\vec\omega)|0\rangle &=& 
- {i\over 2} \,f_P \,\bn\cdot p \,\varphi(u) \, ,\\
\label{scet2p}
\langle P(p) |{\cal P}(\vec\omega) |0\rangle &=&  
-if_P \mu_P\, \varphi_p (u)     \,, \\
\label{scet3p}
 \langle P(p) |\widetilde{\cal P}(\vec\omega) |0\rangle &=& 
           - \frac{i}{6}  \, f_P \mu_P\, \varphi_\sigma^\prime (u)     \,, \\ 
\label{scet4p}
\langle P(p) |{\cal T}^{\alpha \beta}(\vec\omega) |0\rangle &=& 
- \frac{1}{12}   f_P \mu_P \, \epsilon_\perp^{\alpha\beta} \left[ 
  {\bar u - u \over u \bar u}  \varphi_\sigma (u) -
6 R \left( {G_{Px}^{(t)} (u)\over u} + { G_{Py}^{(t)} ( \bar u)\over \bar u} \right) 
\right] \, , \label{tmunu}\\ 
\label{scet5p}
\langle P(p) |\widetilde{\cal T}^{\alpha \beta}(\vec\omega) |0\rangle &=& 
-{1\over 12}f_P \mu_P \,\epsilon_\perp^{\alpha\beta}\left[     
{1\over u \bar u}  \varphi_\sigma (u) -
6R \left(  {G_{Px}^{(t)} (u)\over u} -{ G_{Py}^{(t)} ( u)\over \bar u} \right) 
\right]                       \, , \\
\langle P(p) |{\cal T}_3^{\alpha \beta} (\vec\omega) |0\rangle &=& 
-{1\over 2} f_{3\pi} \, \epsilon_\perp^{\alpha\beta} \,{\varphi_{3\pi} (\vec\alpha) \over \alpha_3^2}\,,
\label{scet3ten} 
\eea 
where $R = f_{3\pi} / (f_P \mu_P)$, $(\omega_1,\omega_2) = (u  , - 
\bar u )\bn \cdot p$, $(\omega_1,\omega_2,\omega_3) = (\alpha_1 , 
-\alpha_2 ,-\alpha_3 )\bn\cdot p$ and 
\bea 
G_{Px}^{(t)} (u) &=& {{\rm d} \over {\rm d}u} \int_0^u {\rm d}\alpha_1 
 \int_0^{\bar u} {\rm d}\alpha_2 \,  {u-\alpha_1 \over \alpha_3^2} 
\varphi_{3\pi} (\vec\alpha) \, ,\\ 
 G_{Py}^{(t)}  (u) &=& {{\rm d} \over {\rm d}u} \int_0^u {\rm d}\alpha_1 
  \int_0^{\bar u} {\rm d}\alpha_2 \,  {\bar u-\alpha_2 \over \alpha_3^2} 
 \varphi_{3\pi} (\vec\alpha)  \, . 
\eea 
All the other matrix elements vanish in the limit in which we neglect
twist-4 wave functions.

Some comments on the manipulations we used to derive the above results 
are necessary. The operator $\widetilde {\cal P}$ does not appear in
Eqs.~(\ref{scs})-(\ref{3te}) but is present in the projection of $\bar
q(x) \sigma^{\mu\nu} \gamma^5 q(y)$, hence it can be expressed in
terms of the above introduced wave functions.  The functions 
$G_{Px}^{(t)} (u)$ and $G_{Py}^{(t)} (u)$ have been obtained by 
inserting Eq.~(\ref{scet3ten}) into Eq.~(\ref{te}), using the 
identities 
\bea 
\int_0^1 {\rm d}t \, t^k \int {\cal D}^3 \vec\alpha \,  {\cal F(\vec\alpha)} 
= 
\int_0^1 {\rm d}u \, e^{i(u p\cdot x + \bar u p\cdot y)} \int_0^u {\rm d}\alpha_1 
\int_0^{\bar u} {\rm d}\alpha_2 \, {1\over \alpha_3} \left[u-\alpha_1\over \alpha_3\right]^k 
      {\cal F(\vec\alpha)}    \, , \\ 
 \int_0^1 {\rm d}t \, {\bar t}^k \int {\cal D}^3 \vec\alpha \,  {\cal F(\vec\alpha)} 
 = 
 \int_0^1 {\rm d}u \, e^{i(u p\cdot x + \bar u p\cdot y)} \int_0^u {\rm d}\alpha_1 
 \int_0^{\bar u} {\rm d}\alpha_2 \, {1\over \alpha_3} \left[\bar u-\alpha_2\over \alpha_3\right]^k 
       {\cal F(\vec\alpha)}    \, ,   \label{62} 
\eea 
which are valid for any function ${\cal F}(\vec \alpha)$ and, finally, 
integrating by parts.  The subscripts $x$ and $y$ indicate that 
$G_{Px}^{(t)} $ and $G_{Py}^{(t)}$ stem from terms proportional to 
$x_\perp$ and $y_\perp$, respectively.  The symmetry property of 
$\varphi_{3\pi} (\vec\alpha)$ under the exchange $\alpha_1 
\leftrightarrow \alpha_2$ implies relations between these two functions. 
For the pion, G-parity implies~\cite{fily2}:
\bea
\varphi_{3\pi} (\alpha_1,\alpha_2,\alpha_3) =  
\varphi_{3\pi} (\alpha_2,\alpha_1,\alpha_3)
& \Longrightarrow &
 G_{Py}^{(t)} ( 1-u)  = -G_{Px}^{(t)} (u) \; . 
\eea 

Finally, the equations of motion in QCD imply relations among the 
twist-3 wave functions. In the approach discussed here they follow
from the relations among SCET operators presented in Sec.~II.
The first such relation is obtained by noting that the matrix element 
of ${\cal T}^{\mu\nu}$ can be directly extracted from the first line in 
Eq.~(\ref{te}). Comparing this determination with Eq.~(\ref{tmunu}) we find
\bea 
\varphi_\sigma^\prime (u) - 
{\bar u - u \over u \bar u} \varphi_\sigma (u) + 6 R \left[ 
{G_{Px}^{(t)} (u)\over u} +{ G_{Py}^{(t)} ( u)\over \bar u} 
\right] = 0   \, . 
\eea 
A second equation of motion follows from using the results (\ref{scet2p}), 
(\ref{scet5p}) in the relation Eq.~(\ref{PtildeT}) and reads
\bea
\phi_p(u) = \frac{1}{6u\bar u}
\phi_\sigma(u) - R\left( \frac{1}{u} G_{Px}^{(t)}(u) - 
\frac{1}{\bar u} G_{Py}^{(t)}(u)\right)\,.
\eea

Neglecting the terms proportional to the 3-parton wave function
$\phi_{3\pi}$, these relations give the so-called Wandzura-Wilczek
relations for the 2-parton twist-3 wave functions \cite{WaWi}. They
can be solved exactly and give the solutions $\phi_p(u)|_{WW} = 1,
\phi_\sigma(u)|_{WW} = 6u\bar u$.

\subsection{Vector meson matrix elements}

Let us collect the matrix elements between the vacuum and a vector
meson state of all the relevant QCD operators~\cite{babr,babr2}.  We
will work in a reference frame where the momentum of the vector meson 
has a large component along the $n$ direction, and a vanishing 
transverse momentum, thus $p_\mu = {\bn\cdot p\over 2} n_\mu$.  We 
neglect a term proportional to the mass squared of the meson, because 
it is irrelevant for the analysis of twist-2 and twist-3 distribution 
amplitudes. 
 
The large light-cone momentum component is given by $\bn\cdot p = 2\, 
E_V + O(m_V^2/E_V)$.  The polarization vector $\eta$ is decomposed as 
a sum of longitudinal $\eta_\parallel$ and transverse $\eta_\perp$ 
components 
\bea 
\eta_\parallel^\mu= {\bn\cdot \eta\over 2} n^\mu 
\,,\qquad \eta_\perp^\mu = \eta^\mu - \eta_\parallel^\mu \, . 
\eea 
Note that these definitions are slightly different from the
conventions used in Refs.~\cite{babr,babr2}; hence, the expressions
for the various matrix elements will look slightly different. Our choice 
of the transverse plane is forced by the structure of the subleading SCET
operators.

The QCD matrix elements read (we often integrate by parts to remove 
$n\cdot z$ factors): 
\bea 
\langle V(p,\eta)|\bar q(x)\, q(y)|0\rangle 
&=&  
-{i\over 2} f_{V}^T \, m_{V}^2 \, z \cdot \eta^* 
\int_0^1 {\rm d}u \,e^{i(u\,p\cdot x+\bar{u}\,p\cdot y)}\, h_{\parallel}^{(s)}(u) \\ 
&=&  
{1\over 2} f_{V}^T \, m_{V}^2 \, \frac{\bar{n} \cdot \eta^*}{\bar {n} \cdot p} 
\int_0^1 {\rm d}u \,e^{i(u\,p\cdot x+\bar{u}\,p\cdot y)}\, {h_{\parallel}^{(s)}}^{\prime}(u) \, , 
\label{vecti} \\ 
\langle V(p,\eta)|\bar q(x)\gamma^5 q(y)|0\rangle 
&=& 
0     \, , 
\eea 
\bea 
\langle V(p,\eta)|\bar q(x)\gamma^{\mu} q(y)|0\rangle 
&=& 
f_{V} \, m_{V} \int_0^1 {\rm d}u \,e^{i(u\,p\cdot x+\bar{u}\,p\cdot y)} \Big(\eta^{\mu *} g_{\perp}^{(v)}(u) 
+ p^\mu {z\cdot \eta^* \over z \cdot p}  (\phi_{\parallel}(u)-g_\perp^{(v)}(u)) \Big) \nn \\ 
&=& 
f_{V} \, m_{V} \int_0^1 {\rm d}u \,e^{i(u\,p\cdot x+\bar{u}\,p\cdot y)} \Big(\eta^{\mu *}_{\perp} g_{\perp}^{(v)}(u) 
+\eta_{\parallel}^{\mu *} \phi_{\parallel}(u) 
 \nn\\ 
& & 
-\frac{i}{2}n^{\mu}\bar{n} \cdot p \, \eta_{\perp}^* \cdot z_{\perp}  F(u)\Big)        \, ,   \\ 
\langle V(p,\eta)|\bar q(x)\gamma^{\mu} \gamma^5 q(y)|0\rangle 
&=& 
- \frac{1}{4} f_{V} \, m_{V} \epsilon^{\mu\nu\rho\sigma} \eta^*_\nu \, p_\rho \, z_\sigma   
  \int_0^1 {\rm d}u \,e^{i(u\,p\cdot x+\bar{u}\,p\cdot y)}\, g_{\perp}^{(a)}(u) \\ 
&=& 
\frac{1}{4} f_{V} \, m_{V} \int_0^1 {\rm d}u \,e^{i(u\,p\cdot x+\bar{u}\,p\cdot y)}\, 
 \Big[i\, \epsilon^{\mu \nu}_{\perp}\eta_{\perp\nu}^{*}\,  {g_{\perp}^{(a)}}^{\prime}(u) 
\nn \\ 
& & 
- \frac{\bar{n} \cdot p}{2} n^{\mu} \epsilon_{\perp}^{\beta \nu} z_{\perp\beta} \eta_{\perp\nu}^{*} 
\, g_{\perp}^{(a)}(u) \Big]         \, , 
\eea 
\bea 
 \langle V(p,\eta)|\bar q(x)\sigma^{\mu \nu} q(y)|0\rangle  
&=& 
- i f_{V}^T \int_0^1 {\rm d}u \,e^{i(u\,p\cdot x+\bar{u}\,p\cdot y)}\, 
\bigg[ (\eta^{\nu *} p^\mu - \eta^{\mu *} p^\nu) \phi_{\perp}(u) \nn \\ 
& & + \left. m_V^2 \frac{z \mcdot \eta^*}{(z \mcdot p)^2} (p^{\mu} z^\nu - p^\nu z^\mu)  
(h_{\parallel}^{(t)}(u) - \phi_\perp(u)) \right] \\
&=& 
- \frac{i}{2} f_{V}^T \int_0^1 {\rm d}u \,e^{i(u\,p\mcdot x+\bar{u}\,p\mcdot y)}\, 
\bigg[ \bar{n} \mcdot p 
\left(\eta_{\perp}^{\mu *}n^{\nu}-n^{\mu}\eta_{\perp}^{\nu *} \right) \phi_{\perp}(u) 
\quad\quad\quad\quad\quad\nn \\ 
& & 
\hskip -3.5cm
+m_V^2 \frac{\bar{n} \mcdot \eta^*}{\bar{n} \mcdot p} 
\left( n^{\mu}\bar{n}^{\nu}-n^{\nu}\bar{n}^{\mu}\right)
h_{\parallel}^{(t)}(u) 
- i m^2 \bn \mcdot \eta^* \left( n^{\mu}z_{\perp}^{\nu}-n^{\nu}z_{\perp}^{\mu} \right) \int_0^u {\rm d}v 
(h_{\parallel}^{(t)}(v) - \phi_\perp(v)) \bigg] \,,\nn
\eea 
\bea 
\langle V(p,\eta)|\bar q(x) g\,G^{\mu \nu} \gamma^{\alpha} q(y)|0\rangle 
&=& 
{i\over 4} f_{V} m_{V} (\bar{n} \mcdot p)^{2} n^{\alpha} 
\left(n^\nu \eta_{\perp}^{\mu *}-n^{\mu} \eta_{\perp}^{\nu *} \right) \int {\cal D}^3  \vec{\alpha} \, 
{\cal V}(\vec\alpha)     \, , 
\\ 
\langle V(p,\eta)|\bar q(x) g\,G^{\mu \nu} \gamma^{\alpha}\gamma^{5}q(y)|0\rangle 
&=& 
\frac{1}{4} f_{V} m_{V} (\bar{n} \mcdot p)^{2} \epsilon^{\mu \nu \rho \sigma} 
n^{\alpha} 
n_{\rho} \eta_{\perp\sigma}^{ *}  \int {\cal D}^3  \vec{\alpha} \, 
{\cal A}(\vec\alpha)      \, ,      \\ 
\langle V(p,\eta)|\bar q(x)\sigma^{\alpha \beta}\, g\,G^{\mu \nu} q(y)|0\rangle 
&=&
\frac{1}{8} f_{V}^T m^{2} \bar{n} \mcdot \eta^* \, \bar{n} \mcdot p 
\Big(n^{\alpha}n^{\mu}g_{\perp}^{\beta \nu}-n^{\beta}n^{\mu}g_{\perp}^{\alpha \nu}-n^{\alpha}n^{\nu} 
g_{\perp}^{\beta \mu}   \quad\quad\quad\quad\nn \\ 
& & 
+n^{\beta}n^{\nu}g_{\perp}^{\alpha \mu} \Big) 
 \int {\cal D}^3 \vec{\alpha}\, {\cal T}(\vec{\alpha})        \, , 
\label{vectf} 
\eea 
where $F(u) = \int_0^u {\rm d} v \, \big[ \phi_\parallel (v) - 
g_\perp^{(v)}(v) \big]$ and $f_V$ and $f_V^T$ are the
vector meson decay constants, defined as
\bea
\langle V(p,\eta)|\bar q\gamma_\mu q|0\rangle = f_V m_V \eta^*_\mu\,,\qquad
\langle V(p,\eta)|\bar q\sigma_{\mu\nu} q|0\rangle = 
-if_V^T (\eta^*_\mu p_\nu - \eta^*_\nu p_\mu)
\eea
In the above formulae, $\phi_\parallel(u)$ and $\phi_\perp(u)$ are of
twist-2, while all the other wave functions are of twist-3. Since
$\eta_\perp / \eta_\parallel \sim m_V/E_V$, we keep terms proportional
to $m^2_V$ only when they involve $\eta_\parallel$.

The extraction of the SCET matrix elements proceeds in complete
analogy with the pseudoscalar meson case. Comparing
Eqs.~(\ref{vecti})-(\ref{vectf}) with Eqs.~(\ref{scs})-(\ref{3te}), we 
obtain: 
\bea 
\langle V(p,\eta) |{\cal J}_V (\vec\omega)|0\rangle &=& 
{1\over 2} \, f_V m_V \, \bn\mcdot \eta^* \, \phi_\parallel (u)\, , 
\quad\quad\quad\quad\quad\quad\quad\quad\quad  \quad 
\quad\quad\quad\quad\quad\quad\quad\quad\quad\quad\\ 
\langle V(p,\eta) |{\cal J}_A (\vec\omega)|0\rangle &=& 0 \, ,\\ 
\langle V(p,\eta) |{\cal J}_T^\mu 
(\vec\omega)|0\rangle &=&  {1\over 4} \,f_V^T \, \eta^{\mu *}_\perp \, \bn\mcdot p \,\phi_\perp(u) \, , 
\eea 
\bea 
\label{v1} 
\langle V(p,\eta) |{\cal V}_1^\mu (\vec\omega) |0\rangle &=&  
f_V m_V \eta_\perp^{\mu *}\, g_\perp^{(v)} (u) \, , 
\quad\quad\quad\quad\quad\quad\quad\quad\quad\quad\quad \quad\quad\quad\quad 
\quad\quad\quad\quad\quad\quad\\ 
\langle V(p,\eta) |\widetilde{\cal V}_1^\mu (\vec\omega) |0\rangle &=&
    {1\over 4} f_V m_V \, \eta_\perp^{\mu *} \, {g_\perp^{(a)}}^\prime (u) \, , \\ 
\langle V(p,\eta) |{\cal V}_2^\mu (\vec\omega) |0\rangle &=& 
-{1\over 2} f_V m_V \eta_\perp^{\mu *} \left[ 
{\bar u - u \over u \bar u} F(u)  -   {G_{Vx}^{(v)} (u)\over u} - { G_{Vy}^{(v)} ( u)\over \bar u} 
\right] 
\, ,   \label{v2} \\ 
\langle V(p,\eta) |\widetilde{\cal V}_2^\mu (\vec\omega) |0\rangle &=& 
- {1\over 2} f_V m_V \eta_\perp^{\mu *}\left[ 
{1 \over u \bar u} F(u)  -   {G_{Vx}^{(v)} (u)\over u} + { G_{Vy}^{(v)} ( u)\over \bar u} 
\right] 
 \, , 
\eea 
\bea 
\langle V(p,\eta) |{\cal A}_1^\mu (\vec\omega) |0\rangle &=& 
{i\over 4 } f_V m_V \, \epsilon_\perp^{\mu\nu} \eta_{\perp\nu}^{ *} \, {g_\perp^{(a)}}^\prime (u) \, , 
\quad\quad\quad\quad\quad\quad\quad\quad\quad\quad\quad\quad\quad\quad\quad\quad\quad\quad\\ 
\langle V(p,\eta) |\widetilde{\cal A}_1^\mu (\vec\omega) |0\rangle &=& 
i f_V m_V \,  \epsilon_\perp^{\mu\nu} \eta_{\perp\nu}^{ *}\, g_\perp^{(v)} (u) 
 \, , \\ 
\langle V(p,\eta) |{\cal A}_2^\mu (\vec\omega) |0\rangle &=& 
{i\over 2} f_V m_V \,  \epsilon_\perp^{\mu\nu}\eta_{\perp\nu}^{*}\left[ 
{\bar u - u \over u \bar u} {g_\perp^{(a)} (u) \over 4} -   {G_{Vx}^{(a)} (u)\over u} - 
 { G_{Vy}^{(a)} (u)\over \bar u} 
\right] 
\, ,\\ 
\langle V(p,\eta) |\widetilde{\cal A}_2^\mu (\vec\omega) |0\rangle &=& 
{i\over 2} f_V m_V \,  \epsilon_\perp^{\mu\nu}\eta_{\perp\nu}^{*}\left[ 
{1 \over u \bar u} {g_\perp^{(a)} (u)\over 4}  -  {G_{Vx}^{(a)} (u)\over u} + 
{ G_{Vy}^{(a)}(u)\over \bar u} 
\right] 
 \, ,    \label{a2tilde} 
\eea 
 \bea 
\langle V(p,\eta)   |{\cal S}(\vec\omega) |0\rangle &=& 
{1\over 2} f_V^T m_V^2 \, {\bn\mcdot \eta^* \over \bn \mcdot p} \,  {h_\parallel^{(s)}}^\prime (u)     \,, 
\quad\quad\quad\quad\quad\quad\quad\quad\quad 
\quad\quad\quad\quad\quad\quad\quad\quad\quad \\ 
 \langle V(p,\eta)  |\widetilde{\cal S}(\vec\omega) |0\rangle &=& 
f_V^T m_V^2 \, {\bn\mcdot \eta^* \over \bn \mcdot p} \,  {h_\parallel^{(t)}} (u) 
              \,,     \\ 
\langle V(p,\eta)   |{\cal P}(\vec\omega) |0\rangle &=& 0    \,, \\ 
 \langle V(p,\eta)  |\widetilde{\cal P}(\vec\omega) |0\rangle &=&        0 
               \,, 
\eea 
\bea 
\langle V(p,\eta) |{\cal T}^{\mu\nu}(\vec\omega) |0\rangle &=& 
 {1\over 2} f_V^T m_V^2 \, {\bn \mcdot \eta^* \over \bn \mcdot p} \, g_\perp^{\mu\nu} \,\\
& & \times \left[
{\bar u - u \over u \bar u}     \int_0^u {\rm d} v \, (h_\parallel^{(t)}(v) - \phi_\perp(v))
+ {G_{Vx}^{(t)} (u)\over 2 u} + { G_{Vy}^{(t)} (u)\over 2 \bar u} 
\right] \,,\nn             \\ 
\langle V(p,\eta) |\widetilde {\cal T}^{\mu\nu}(\vec\omega) |0\rangle &=& 
{1\over 2} f_V^T m_V^2 \, {\bn \mcdot \eta^* \over \bn \mcdot p} \, g_\perp^{\mu\nu} \,\\
& &\times \left[
 {1 \over u \bar u} \int_0^u {\rm d} v \, (h_\parallel^{(t)}(v)-\phi_\perp(v))
 + {G_{Vx}^{(t)} (u)\over 2 u} - { G_{Vy}^{(t)} (u)\over 2 \bar u}
\right] \,,\nn
\eea 
\bea 
\langle V(p,\eta) |{\cal V}_3^{\mu} (\vec\omega) |0\rangle &=& 
{1\over 2} f_V m_V \, \eta_\perp^{\mu *} \, {{\cal V} (\vec\alpha) \over \alpha_3^2}        \,, 
\quad\quad\quad\quad\quad\quad\quad\quad\quad\quad\quad\quad\quad\quad\quad\quad 
\quad\quad\quad\quad\quad  \\ 
\langle V(p,\eta) |{\cal A}_3^{\mu} (\vec\omega) |0\rangle &=& 
-{i\over 2} f_V m_V \, \epsilon_\perp^{\mu\nu}\eta_{\perp\nu}^{*} \, {{\cal A} (\vec\alpha) \over \alpha_3^2}        \,, \\ 
\langle V(p,\eta) |{\cal T}_3^{\mu\nu} (\vec\omega) |0\rangle &=& 
 {1\over 4} f_V^T m_V^2 \, {\bn \mcdot \eta^* \over \bn \mcdot p} \, g_\perp^{\mu\nu} \, {{\cal T} (\vec\alpha) \over \alpha_3^2} 
\, , 
\eea 
where 
 \bea 
G_{Vx}^{(v,a,t)} (u) = {{\rm d} \over {\rm d}u} \int_0^u {\rm d}\alpha_1 
 \int_0^{\bar u} {\rm d}\alpha_2 \,  {u-\alpha_1 \over \alpha_3^2} 
({\cal V},{\cal A},{\cal T}) (\vec\alpha) \, ,    \label{3parx}   \\ 
 G_{Vy}^{(v,a,t)} (u) = {{\rm d} \over {\rm d}u} \int_0^u {\rm d}\alpha_1 
  \int_0^{\bar u} {\rm d}\alpha_2 \,  {\bar u-\alpha_2 \over \alpha_3^2} 
 ({\cal V},{\cal A},{\cal T}) (\vec\alpha) \, .    \label{3pary} 
\eea 
For the case of the $\rho$ meson in the limit of isospin symmetry,
G-parity requires that the 3-parton wave functions satisfy 
${\cal A}(\alpha_1,\alpha_2,\alpha_3) =
{\cal A}(\alpha_2,\alpha_1,\alpha_3)$  and $\{{\cal V}, {\cal T}\}
(\alpha_1,\alpha_2,\alpha_3) = -\{{\cal V}, {\cal T}\}
(\alpha_2,\alpha_1,\alpha_3)$. 
This implies the relations $G_{Vy}^{(a)}(\bar u) = -G_{Vx}^{(a)}(u)$
and $G_{Vy}^{(v,t)}(\bar u) = G_{Vx}^{(v,t)}(u)$.

Note that $\widetilde{\cal V}_1 $ and $\widetilde{\cal A}_1 $ do not
appear in Eqs.~(\ref{scs})-(\ref{3te}) but their matrix elements are
easily obtained via Eq.~(\ref{tilde}). The computation of all the
other matrix elements is similar to the pseudoscalar meson case.

The insertion of the matrix elements, Eqs.~(\ref{v1})-(\ref{a2tilde}), 
into Eq.~(\ref{ww}) results in the following relations between the 
chiral-even light-cone wave functions:
\bea 
\label{ww1} 
 2 \, g_\perp^{(v)}(u)  - \frac{1}{4} \, \frac{g_\perp^{(a)}(u)}{u\bar u}  +  
\frac{\bar u-u}{u\bar u} \, F(u) 
&= & 
 \frac{G_{Vx}^{(v)} (u) - G_{Vx}^{(a)} (u)}{u}+\frac{G_{Vy}^{(v)} (u) + 
G_{Vy}^{(a)} (u)}{\bar u} \, ,\\
\label{ww2}
\frac{1}{2}\, g^{\prime(a)}_\perp(u) - \frac{1}{4}\, \frac{\bar u - u}{u\bar u} \, 
g_\perp^{(a)}(u)  +  \frac{F(u)}{u\bar u} 
&=& 
 \frac{G_{Vx}^{(v)} (u) - G_{Vx}^{(a)} (u)}{u}- \frac{G_{Vy}^{(v)} (u) + 
G_{Vy}^{(a)} (u)}{\bar u}  \,.
\eea 
The identities Eq.~(\ref{StildeT}) imply two other relations for the
chiral-odd wave functions
\bea
h_\parallel^{(t)}(u) - \frac{\bar u - u}{u\bar u} 
\int_0^u {\rm d}v (h_\parallel^{(t)}(v) - \phi_\perp(v)) &=& 
\frac{1}{2u} G_{Vx}^{(t)}(u) + \frac{1}{2\bar u} G_{Vy}^{(t)}(u)\,,\\
\frac{d}{du} h_\parallel^{(s)}(u) - \frac{2}{u\bar u}
\int_0^u {\rm d}v (h_\parallel^{(t)}(v) - \phi_\perp(v)) &=&
\frac{1}{u} G_{Vx}^{(t)}(u) - \frac{1}{\bar u} G_{Vy}^{(t)}(u)\,.
\eea

In the limit in which we neglect higher Fock state contributions (i.e.
where we set to zero the right-hand sides of
Eqs.~(\ref{ww1})-(\ref{ww2})), we obtain Wandzura-Wilczek like
relations between twist-2 and twist-3 wave functions~\cite{WaWi}. They
can be again solved exactly \cite{babr,babr2} and give the twist-3
wave functions in terms of the twist-2 ones $\phi_\parallel(u),
\phi_\perp(u)$. For the chiral-even structures one finds
\bea
g_\perp^{(v)}(u)\Big|_{\rm WW} &=& 
\frac12 \left[ \int_0^u {\rm d}v \frac{\phi_\parallel(v)}{\bar v}
+ \int_u^1 {\rm d}v \frac{\phi_\parallel(v)}{v} \right]\\
g_\perp^{(a)}(u)\Big|_{\rm WW} &=& 
2 \left[ \bar u \int_0^u {\rm d}v \frac{\phi_\parallel(v)}{\bar v}
+ u \int_u^1 {\rm d}v \frac{\phi_\parallel(v)}{v} \right]\,,
\eea
and for the chiral-odd wave functions
\bea
h_\parallel^{(t)}(u)\Big|_{\rm WW} &=& 
(2u-1) \left[ \int_0^u {\rm d}v \frac{\phi_\perp(v)}{\bar v}
- \int_u^1 {\rm d}v \frac{\phi_\perp(v)}{v} \right]\\
h_\parallel^{(s)}(u)\Big|_{\rm WW} &=& 
2 \left[ \bar u \int_0^u {\rm d}v \frac{\phi_\perp(v)}{\bar v}
+ u \int_u^1 {\rm d}v \frac{\phi_\perp(v)}{v} \right]\,.
\eea

\section{Reparameterization invariance constraints}
\label{RPI}

The soft-collinear effective theory has an additional symmetry,
related to the Lorentz invariance of the full theory, which was
explicitly broken by defining the effective theory in terms of the
arbitrary light-cone vectors $n_\mu$ and $\bn_\mu$. This symmetry
manifests itself as an invariance under small changes in the
light-cone vectors $n_\mu$ and $\bar n_\mu$, and is usually called
reparameterization invariance (RPI).  The RPI has been shown to impose
rather stringent constraints on the form of the effective theory
Lagrangian and heavy-light currents \cite{ChayKim,MMPS,BCDF,HN,ps1}.
For earlier uses of Lorentz invariance in hard scattering processes,
see Refs.~\cite{Lorentz}.

In this Section we show that reparameterization invariance can be used
to derive strict constraints on the Wilson coefficients of the
subleading collinear operators considered in this paper. In
particular, these constraints fix the coefficients of the
three-parton operators of the form $\bar\chi_{n,\omega_1} [W^\dagger
iD_\perp W]_{\omega_3} \chi_{n,\omega_2}$, in terms of coefficients of
the leading two-body operators of type $\bar \chi_{n,\omega_1} \Gamma
\chi_{n,\omega_2}$.

The explicit form of the RPI constraints for the subleading collinear
operators is in general process dependent, and depends on the SCET
operators which are allowed in the expansion of the physical quantity
being described (effective Hamiltonian, current, etc.).  Despite this
diversity, there is one general feature which is common to all these
situations: the SCET expansion must contain (at least) one additional
vector $z_\mu$, in addition to $n_\mu$ and $\bar n_\mu$.  This vector can be
for example $v_\mu$, the heavy quark velocity in problems involving
heavy quark decay, or a space-time vector $z_\mu$, describing the
nonlocality of a T-product, as for example in hard scattering
processes.  The presence of the additional vector is required by type
III RPI: the Wilson coefficients of the SCET operators must depend on
RPI invariant arguments, and it is impossible to form such
combinations from $\bn\mcdot {\cal P}$ and $\bn\cdot {\cal P}^\dagger$
alone. The situation is different in the presence of an additional
vector $z$, when RPI-III invariant combinations can be formed as
$n\cdot z \bn\cdot {\cal P}$ and $n\cdot z \bn\cdot {\cal P}^\dagger$.

In the following we illustrate the form of the RPI constraints using 
a few examples. We will derive the RPI constraints on the Wilson 
coefficients in the matching of operators with scalar and vector quantum 
numbers, which will be left completely unspecified.

\subsection{Scalar operator}

Consider first a scalar QCD chiral-even operator $S(z)$, depending on 
an arbitrary vector $z_\mu$. Keeping terms of $O(1,\lambda)$, the most 
general form
for this operator in SCET contains the terms (integration over
$\omega_i$ is implied on the RHS)
\bea\label{sexp}
S(z) &=&  C(\omega_1,\omega_2)(n\mcdot z) {\cal J}_V(\omega_1,\omega_2) +
\sum_{i=1}^2 D_i(\omega_1,\omega_2) (z_\alpha {\cal V}_i^\alpha)\\
&+&  \sum_{i=1}^2 \tilde D_i(\omega_1,\omega_2) 
(z_\alpha \tilde {\cal V}_i^\alpha) 
+  E(\omega_1,\omega_2,\omega_3) (z_\alpha {\cal V}_3^\alpha)\,.\nn
\eea
This expansion contains the most general structures transforming as a scalar
under Lorentz transformations, and which are chiral-even and invariant under 
type-III RPI. 
As mentioned above, the Wilson coefficients $C,D_i,\tilde D,
E$ are functions of the type III invariants 
$n\cdot z \omega_1$ and $n\cdot z \omega_2$.
In the presence of more than one vector $z$, the possible  number
of structures on the right-hand side is correspondingly larger. 
The Wilson coefficients will also depend on
all possible type III RPI invariant combinations $n\cdot z_i \bn\cdot
{\cal P}^{(\dagger)}$. The constraints derived below can be
straightforwardly generalised to such cases.

We will show in the following that type-I and -II RPI gives stringent
constraints on the form of the Wilson coefficients appearing in 
Eq.~(\ref{sexp}). There are two constraints following from type-I
RPI
\bea\label{I-1s}
\mbox{(I-1): } & & 
\left( 1 + \omega_1 \frac{\partial}{\partial \omega_1} +
\omega_2 \frac{\partial}{\partial\omega_2}\right)
C(\omega_1,\omega_2) - \sum_{i=1,2}D_i(\omega_1,\omega_2)
-\frac12 E(\omega_1,\omega_2,0) = 0\,, \\
\label{I-2s}
\mbox{(I-2): } & & \tilde D_1(\omega_1,\omega_2) = 0\,.
\eea
and five other constraints from type-II RPI
\bea\label{II-1s}
\mbox{(II-1): } & & C(\omega_1,\omega_2) - D_1(\omega_1,\omega_2) = 0\,, \\
\label{II-2s}
\mbox{(II-2): } & & \left( \omega_1 \frac{\partial}{\partial \omega_1} +
\omega_2 \frac{\partial}{\partial\omega_2}\right)
C(\omega_1,\omega_2) - D_2(\omega_1,\omega_2) = 0\,, \\
\label{II-3s}
\mbox{(II-3): } 
& & \tilde D_1(\omega_1,\omega_2) = 0\,, \\
\label{II-4s}
\mbox{(II-4): } 
& & \left( \omega_1 \frac{\partial}{\partial \omega_1} -
\omega_2 \frac{\partial}{\partial\omega_2}\right)
C(\omega_1,\omega_2) - \tilde D_2(\omega_1,\omega_2) = 0\,, \\
\label{II-5s}
\mbox{(II-5): } & & 
C(\omega_1-\omega_3,\omega_2) - C(\omega_1,\omega_2+\omega_3)\\
& & +  \omega_3 
\left( \frac{\partial}{\partial \omega_1} C(\omega_1-\omega_3,\omega_2) +
\frac{\partial}{\partial\omega_2}C(\omega_1,\omega_2+\omega_3) \right)\nn\\
& & + \frac12 E(\omega_1,\omega_2,\omega_3) = 0\,.\nn
\eea

In the following we present the derivation of these constraints, focusing
on the type-II RPI constraints, which are technically more involved.  
We start by
computing the variation of the leading $O(\lambda^0)$ term in
(\ref{sexp}) under a type II reparameterization transformation $\bn_\mu
\to \bn_\mu + \varepsilon^\perp_\mu$.  This receives contributions
from the change in the Wilson coefficient and in the operator 
%
\bea\label{start}
\delta_{\rm II} [C(n\mcdot z \omega_i) {\cal J}_V(\omega_i)] =
\delta_{\rm II} [C(n\mcdot z \omega_i)] {\cal J}_V(\omega_i) +
C(n\mcdot z \omega_i) \delta_{\rm II} [{\cal J}_V(\omega_i)]\,.
\eea
The variation of the leading order operator ${\cal J}_V(\omega_i)$ in
Eq.~(\ref{start}) can be computed using the action of the RP
transformation on the collinear fields given in \cite{MMPS}. One finds
\bea
\delta^{(\lambda)}_{\rm II} {\cal J}_V(\omega_i) &=&
\left[ \bar\xi_n \frac{\bnslash}{2} (i\Dslash_{c\perp})^\dagger W\right]_{\omega_1}
\frac{1}{\bn\mcdot {\cal P}^\dagger} \frac{\epsslash_\perp}{2} \chi_{n,\omega_2} +
\bar\chi_{n,\omega_1} \frac{\epsslash_\perp}{2} \frac{1}{\bn\mcdot {\cal P}}
\left[ W^\dagger i\Dslash_{c\perp} \frac{\bnslash}{2} \xi_n\right]_{\omega_2}\\
& &-\int d\omega_3 \left\{ \bar\chi_{n,\omega_1+\omega_3} 
\left[ \frac{1}{\bn\mcdot {\cal P}}
W^\dagger \varepsilon_\perp \mcdot iD_{c\perp} W\right]_{\omega_3} \frac{\bnslash}{2}
\chi_{n,\omega_2} \right.\nn\\ 
& & \left. + \bar\chi_{n,\omega_1} \frac{\bnslash}{2} \left[
W^\dagger (\varepsilon_\perp \mcdot iD_{c\perp})^\dagger W \frac{1}{\bn\mcdot {\cal P}^\dagger}
\right]_{\omega_3} \chi_{n,\omega_2-\omega_3}\right\} \nn\\
&=&  \frac12 \varepsilon_{\perp}\mcdot {\cal V}_1(\omega_i)
- \int {\rm d}\omega_3 \left\{
\varepsilon_{\perp}\mcdot {\cal V}_3(\omega_1+\omega_3,\omega_2,\omega_3) 
- \varepsilon_{\perp}\mcdot {\cal V}_3(\omega_1,\omega_2-\omega_3,\omega_3)\right\}\,,
\nn
\eea
where the first line comes from the change in the collinear quark
fields, and the second from the variation of $W$. 

The variation of the Wilson coefficient in Eq.~(\ref{start}) can be found 
by recalling that
$\omega_1 \to \bn\cdot {\cal P}^\dagger$ and $\omega_2 \to \bn\cdot {\cal P}$
\bea
\delta_{\rm II} [C(n\mcdot z \omega_i)] {\cal J}_V 
&=& \frac{\partial C(n\mcdot z \omega_i)}{\partial \omega_1} 
[\bar\chi_{\omega_1} 
(\varepsilon_\perp\mcdot {\cal P}^\dagger_\perp) 
\frac{\bnslash}{2} \chi_{\omega_2}] +
\frac{\partial C(n\mcdot z \omega_i) }{\partial \omega_2} 
 [\bar\chi_{\omega_1} 
\frac{\bnslash}{2}(\varepsilon_\perp\mcdot {\cal P}_\perp) \chi_{\omega_2}]
\,.\nn\\
\eea
The two terms can further be written in terms of the operators
introduced in Sec.~\ref{bilinears} as
\bea
\bar\chi_{\omega_1} (\varepsilon_\perp\mcdot {\cal P}^\dagger_\perp) 
\frac{\bnslash}{2} \chi_{\omega_2} &=& \left[
\bar\xi_n \varepsilon\mcdot (iD_{c\perp})^\dagger W\right]_{\omega_1} 
\frac{\bnslash}{2} \chi_{\omega_2} - 
[\bar\xi_n \varepsilon\mcdot iD_{c\perp} W]_{\omega_1} \frac{\bnslash}{2}
\chi_{\omega_2}\\
& = & \frac12 \omega_1 [\varepsilon_{\perp}\mcdot {\cal V}_2 + 
\varepsilon_{\perp}\mcdot \tilde {\cal V}_2 ]
- \int {\rm d}\omega_3 \omega_3
\varepsilon{_\perp}\mcdot 
{\cal V}_3(\omega_1+\omega_3,\omega_2,\omega_3)\nonumber
\eea
and
\bea
\bar\chi_{\omega_1} \frac{\bnslash}{2} (\varepsilon_\perp\mcdot {\cal P}_\perp) 
\chi_{\omega_2} &=&
\bar\chi_{\omega_1} \frac{\bnslash}{2} [W^\dagger \varepsilon_\perp \mcdot iD_{c\perp}
\xi_n]_{\omega_2} - \bar\chi_{\omega_1} \frac{\bnslash}{2} 
[W^\dagger \varepsilon_\perp\mcdot (iD_{c\perp})^\dagger \xi_n]_{\omega_2} \\
& =&\frac12 \omega_2 [\varepsilon_\perp\mcdot {\cal V}_2 - 
\varepsilon_{\perp}\mcdot \tilde {\cal V}_2]
- \int {\rm d}\omega_3 \omega_3 
\varepsilon_{\perp}\mcdot {\cal V}_3 (\omega_1,\omega_2-\omega_3,\omega_3) \,.
\nonumber
\eea
Combining everything, one finds the following total result for the 
variation of the leading term in Eq.~(\ref{sexp}) under a type-II RPI
\bea\label{delII}
\delta^{(\lambda)}_{\rm II} [C(n\mcdot z \omega_i)  {\cal J}_V(\omega_i)] &=& \\
& & \hskip -2.2cm
\left[\frac{\partial}{\partial \omega_1}C(\omega_i)\right] 
\left[
\frac12 \omega_1 [\varepsilon_\perp\mcdot {\cal V}_2 + \varepsilon_\perp\mcdot \tilde {\cal V}_2]
- \int d\omega_3 \omega_3
\varepsilon_\perp\mcdot {\cal V}_3(\omega_1+\omega_3,\omega_2,\omega_3)\right] 
\nonumber\\
&&  \hskip -2.2cm
+ \left[\frac{\partial}{\partial \omega_2}C(\omega_i)\right] 
\left[\frac12 \omega_2 [\varepsilon_\perp\mcdot {\cal V}_2 - 
\varepsilon_\perp\mcdot \tilde {\cal V}_2]
- \int d\omega_3 \omega_3
\varepsilon_\perp\mcdot {\cal V}_3(\omega_1,\omega_2-\omega_3,\omega_3)\right] \nonumber\\
&&  \hskip -2.2cm
+C(\omega_i)  \left[
\frac12 \varepsilon_\perp\mcdot {\cal V}_1(\omega_i)
- \int d\omega_3 \left( \varepsilon_\perp\mcdot {\cal
V}_3(\omega_1+\omega_3,\omega_2,\omega_3) 
- \varepsilon_\perp\mcdot
{\cal V}_3(\omega_1,\omega_2-\omega_3,\omega_3)\right)\right] \,.\nonumber
\eea

This has to be cancelled by the $O(\lambda)$ terms in the variation of
the power suppressed terms in Eq.~(\ref{sexp}).  The transformation of
all collinear subleading operators ${\cal V}_{1,2},\tilde {\cal
  V}_{1,2}$ and ${\cal V}_3$ is the same, and is given by
\bea
\delta_{\rm II}^{(\lambda)} {\cal V}_i^\alpha(\omega_i) = -\frac12 n^\alpha 
\varepsilon_\perp\mcdot {\cal V}_i(\omega_i)\,.
\eea
Using this relation in Eq.~(\ref{sexp}), and requiring the
cancellation of all independent structures, gives the type-II RPI
constraints Eqs.~(\ref{II-1s})-(\ref{II-5s}).

The type-I RPI constraints Eqs.~(\ref{I-1s})-(\ref{I-2s}) are derived in
a very similar way, so we only sketch the relevant steps.  Under a
type I RPI transformation $n_\mu \to n_\mu + \Delta^\perp_\mu$, the
variation of the leading order term in Eq.~(\ref{sexp}) is given by
\bea
\delta_I [ C(\omega_i) {\cal J}_V(\omega_i)] &=& 
 (\Delta_\perp\mcdot z_\perp) \left[\left(
\omega_1 \frac{\partial}{\partial \omega_1} + \omega_2 \frac{\partial}{\partial \omega_2}
\right) C(\omega_i)\right] {\cal J}_V(\omega_i)\,.
\eea
This has to cancel the $O(\lambda)$ term in the variation of the
subleading terms.  The action of the type I RPI transformation on the
individual operators is easily obtained as
\bea
& &\delta_I^{(\lambda)} {\cal V}_1^\mu(\omega_i) = 
\delta_I^{(\lambda)} {\cal V}_2^\mu(\omega_i) = -\Delta^\perp_\mu {\cal J}_V(\omega_i) \,, \\
& &\delta_I^{(\lambda)} \tilde {\cal V}_1^\mu(\omega_i) =
-i\varepsilon_\perp^{\mu\nu} \Delta^\perp_\nu {\cal J}_A(\omega_i) \,, \\
& &\delta_I^{(\lambda)} \tilde {\cal V}_2^\mu(\omega_i) = 0  \,, \\
& &\delta_I^{(\lambda)} {\cal V}_3^\mu(\omega_1\omega_2\omega_3) =
-\frac12 \Delta_\perp^\mu {\cal J}_V(\omega_1\omega_2) \delta(\omega_3)\,.
\eea
Inserting this into Eq.~(\ref{sexp}), it is easy to see that the total
variation of the $\tilde {\cal V}_1^\mu$ terms cannot be cancelled by
anything else. This gives $\tilde D_1 = 0$. Requiring the
cancellation of the terms proportional to $(z_\perp\cdot \Delta^\perp) 
{\cal J}_V$ gives Eq.~(\ref{I-1}).
This completes the proof of the RPI constraints.

We note here the remarkable result that the Wilson coefficients of the 
$O(\lambda)$ operators in the expansion of a scalar operator
are completely fixed by RPI in terms of the
Wilson coefficient $C(\omega_i)$ of the leading operator. 
This is different from the case for the heavy-light currents studied in
Ref.~\cite{ps1}, where several of the subleading operators were not
constrained from symmetry arguments alone, and their Wilson coefficients
have to be determined by explicit matching computations.

\subsection{Vector operator}

We consider next the reparameterization invariance constraints on the
Wilson coefficients in the matching of a chiral-even operator
with the quantum numbers of the vector current $V_\mu(z)$. For this case,
the  SCET expansion can contain more terms. Keeping again terms up to
subleading order we can write again the most general type-III RP
invariant structure as
\bea\label{vexp}
V_\mu(z) &=& n_\mu C(\omega_1,\omega_2) {\cal J}_V(\omega_1,\omega_2) +
\sum_{i=1}^2 B_i(\omega_1,\omega_2) {\cal V}_{i\mu} + 
\sum_{i=1}^2 \tilde B_i(\omega_1,\omega_2) \tilde {\cal V}_{i\mu} \\
&+& \frac{n_\mu}{n\mcdot z} \sum_{i=1}^2 D_i(\omega_1,\omega_2) 
(z\mcdot {\cal V}_{i}) + 
\frac{n_\mu}{n\mcdot z} \sum_{i=1}^2 \tilde D_i(\omega_1,\omega_2) 
(z\mcdot \tilde {\cal V}_{i})\nn\\
&+&  E_1(\omega_1,\omega_2,\omega_3) {\cal V}_{3\mu} + 
\frac{n_\mu}{n\mcdot z}  E_2(\omega_1,\omega_2,\omega_3) 
(z\mcdot {\cal V}_{3})\nn\,.
\eea
We neglect here terms of the form $z_\mu S(z)$ with $S(z)$ a scalar
quantity which is RP invariant by itself. The most general form for such
a term has been discussed in the previous section.

The constraints for this case can be derived in analogy with those for
the scalar current. We find four constraints from type I RPI
\bea\label{I-1}
\mbox{(I-1): } & & C(\omega_1,\omega_2) - \sum_{i=1,2}B_i(\omega_1,\omega_2) -
\frac12 E_1(\omega_1,\omega_2,0) = 0\,, \\
\label{I-2}
\mbox{(I-2): } & & \left( \omega_1 \frac{\partial}{\partial \omega_1} +
\omega_2 \frac{\partial}{\partial\omega_2}\right)
C(\omega_1,\omega_2) - \sum_{i=1,2}D_i(\omega_1,\omega_2) -
\frac12 E_2(\omega_1,\omega_2,0) = 0\,, \\
\label{I-3}
\mbox{(I-3): } & & \tilde B_1(\omega_1,\omega_2) = 0\,, \\
\label{I-4}
\mbox{(I-4): } & & \tilde D_1(\omega_1,\omega_2) = 0\,.
\eea
and four other constraints from type II RPI
\bea\label{II-1}
\mbox{(II-1): } & & C(\omega_1,\omega_2) - B_1(\omega_1,\omega_2) -
D_1(\omega_1,\omega_2) = 0\,, \\
\label{II-2}
\mbox{(II-2): } & & \left( \omega_1 \frac{\partial}{\partial \omega_1} +
\omega_2 \frac{\partial}{\partial\omega_2}\right)
C(\omega_1,\omega_2) - B_2(\omega_1,\omega_2) -
D_2(\omega_1,\omega_2) = 0\,, \\
\label{II-3}
\mbox{(II-3): } & & \left( \omega_1 \frac{\partial}{\partial \omega_1} -
\omega_2 \frac{\partial}{\partial\omega_2}\right)
C(\omega_1,\omega_2) - \tilde B_2(\omega_1,\omega_2) -
\tilde D_2(\omega_1,\omega_2) = 0\,, \\
\label{II-4}
\mbox{(II-4): } & & C(\omega_1-\omega_3,\omega_2) - C(\omega_1,\omega_2+\omega_3)\\
& & + \omega_3 
\left( \frac{\partial}{\partial \omega_1} C(\omega_1-\omega_3,\omega_2) +
\frac{\partial}{\partial\omega_2}C(\omega_1,\omega_2+\omega_3) \right)\nn\\
& & + \frac12 \sum_{i=1,2} E_i(\omega_1,\omega_2,\omega_3) = 0\,.\nn
\eea

It is instructive to compare the RPI constraints Eqs.~(\ref{I-1})-(\ref{II-4}) 
with the explicit results for the matching of the nonlocal vector operator
$\bar q(x) \gamma_\mu q(y)$ onto SCET operators given in Eq.~(\ref{scv}). 
For simplicity we take $x_\mu = -y_\mu = z_\mu$ in Eq.~(\ref{scv}), which can
then be written in a form similar to Eq.~(\ref{vexp}) provided that one takes
\bea
C(\omega_1,\omega_2) &=& e^{\frac{i}{2}n.z(\omega_1 + \omega_2)}\,, \\
B_1(\omega_1,\omega_2) &=& C(\omega_i)\,, \quad B_2(\omega_i) = 0\,,\quad 
\tilde B_{1,2}(\omega_i) = 0\,, \\
D_1(\omega_1,\omega_2) &=& 0\,,\quad D_2(\omega_i) = 
\frac{i}{2}n\mcdot z (\omega_1 + \omega_2)C(\omega_1,\omega_2)\,, \\
\tilde D_1(\omega_1,\omega_2) &=& 0\,,\quad \tilde D_2(\omega_i) = 
\frac{i}{2}n\mcdot z (\omega_1 - \omega_2) C(\omega_1,\omega_2)\,, \\
E_1(\omega_1,\omega_2,\omega_3) &=& 0\,,\quad E_2(\omega_i) = 
-\int_0^1 dt (2t-1) \omega_3^2 (n\mcdot z)^2
e^{\frac{i}{2}n.z(\omega_1 + \omega_2 + \omega_3(2t-1))} \,.
\eea
Performing the integration in $E_2(\omega_i)$ one finds
\bea
E_2(\omega_i) = -2\left(1 - \frac{i}{2}n\cdot z \omega_3 \right)
e^{\frac{i}{2}n.z(\omega_1 + \omega_2 + \omega_3)} +
2\left(1 + \frac{i}{2}n\cdot z \omega_3 \right)
e^{\frac{i}{2}n.z(\omega_1 + \omega_2 - \omega_3)}\,.
\eea

It is easy to see that the constraints Eqs.~(\ref{I-1})-(\ref{II-4})
are indeed satisfied with these results for the Wilson coefficients.
For the case of the nonlocal vector current, the matching
Eq.~(\ref{scv}) can be worked out using the equations of motion in QCD
and is exact to all orders. The true power of the RPI constraints
Eqs.~(\ref{I-1})-(\ref{II-4}) becomes apparent in those cases
where the Wilson coefficients have to be obtained in perturbation
theory, and where such constraints can provide a useful check.

\section{Application: weak annihilation in $B\to V\gamma$ decays}
\label{ward}

In this Section we discuss an application where the subleading
operators constructed in Sec.~\ref{matrixelements} appear in a 
situation of physical interest. Consider
real photon emission from a energetic $q\bar q$ pair, which subsequently
hadronizes into a meson. This process is relevant for the
rare radiative decay $B\to V\gamma$, where it contributes to the
weak annihilation amplitude.

In the following we will match the amplitudes contributing to this
process onto SCET operators, including the three-parton operators
constructed in Sec.~\ref{bilinears}. The matrix elements of these
operators are then computed with the help of the relations in 
Sec.~\ref{matrixelements}. For the case of a vector current, the
corresponding matrix element is fixed by an exact Ward identity 
\cite{GrPi,KhWy}.
In addition to serving as illustration for the use of the matrix 
elements computed here, this will provide, at the same time,
a strong check for the consistency of our results.

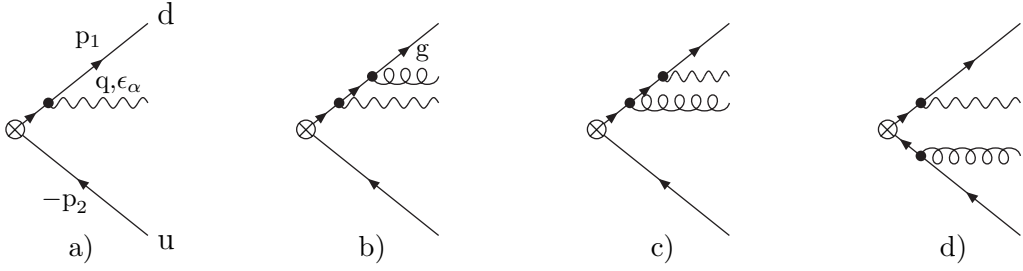
\begin{figure}[t] 
\begin{center}
\begin{picture}(50,100)(170,0)
\ArrowLine(0,50)(12.5,60)
\ArrowLine(12.5,60)(50,90)
\ArrowLine(50,10)(0,50)
\Vertex(12.5,60){2}
\Photon(12.5,60)(50,60){-2}{5}
\BCirc(0,50){3.5}
\Text(50,90)[bl]{\begin{normalsize} ${\rm d}$ \end{normalsize}}
\Text(50,10)[tl]{\begin{normalsize} ${\rm u}$ \end{normalsize}}
\Text(31,27)[tr]{\begin{footnotesize} {$ {\rm -p}_{2}$} \end{footnotesize}}
\Text(36,81)[br]{\begin{footnotesize} {$ {\rm p}_{1}$} \end{footnotesize}}
\Text(27,64)[bl]{\begin{footnotesize} {$ {\rm q},\! \epsilon_{\alpha}$} \end{footnotesize}}
\Text(0,50)[c]{$\times$}
\Text(25,5)[c]{${\rm a)}$}
\SetOffset(110,0)
\ArrowLine(0,50)(12.5,60)
\ArrowLine(12.5,60)(25,70)
\ArrowLine(25,70)(50,90)
\ArrowLine(50,10)(0,50)
\Vertex(12.5,60){2}
\Vertex(25,70){2}
\Photon(12.5,60)(50,60){-2}{5}
\Gluon(25,70)(50,70){-3}{3}
\BCirc(0,50){3.5}
\Text(50,76)[br]{\begin{footnotesize} {$ {\rm g}$} \end{footnotesize}}
\Text(0,50)[c]{$\times$}
\Text(25,5)[c]{${\rm b)}$}
\SetOffset(220,0)
\ArrowLine(0,50)(12.5,60)
\ArrowLine(12.5,60)(25,70)
\ArrowLine(25,70)(50,90)
\ArrowLine(50,10)(0,50)
\Vertex(12.5,60){2}
\Vertex(25,70){2}
\Gluon(12.5,60)(50,60){-3}{5}
\Photon(25,70)(50,70){-2}{4}
\BCirc(0,50){3.5}
\Text(0,50)[c]{$\times$}
\Text(25,5)[c]{${\rm c)}$}
\SetOffset(330,0)
\Vertex(0,50){2}
\ArrowLine(0,50)(12.5,60)
\ArrowLine(12.5,60)(50,90)
\ArrowLine(50,10)(12.5,40)
\ArrowLine(12.5,40)(0,50)
\Vertex(12.5,60){2}
\Vertex(12.5,40){2}
\Gluon(12.5,40)(50,40){3}{5}
\Photon(12.5,60)(50,60){-2}{5}
\BCirc(0,50){3.5}
\Text(0,50)[c]{$\times$}
\Text(25,5)[c]{${\rm d)}$}
\end{picture}
\end{center} 
\caption{QCD diagrams contributing to photon emission from a $q\bar q$
pair created by the weak vertex. The cross indicates an insertion of the 
weak current $\gamma^\mu {1-\gamma^5 \over 2}$.} 
\end{figure} 
\begin{figure}[t] 
\begin{center}
\begin{picture}(50,100)(170,0)
\DashArrowLine(0,50)(50,90){4}
\DashArrowLine(50,10)(0,50){4}
\Photon(0,50)(50,50){2}{7}
\BCirc(0,50){3.5}
\Text(50,90)[bl]{\begin{normalsize} ${\rm d}$ \end{normalsize}}
\Text(50,10)[tl]{\begin{normalsize} ${\rm u}$ \end{normalsize}}
\Text(31,27)[tr]{\begin{footnotesize} {$ {\rm -p}_{2}$} \end{footnotesize}}
\Text(30,75)[br]{\begin{footnotesize} {$ {\rm p}_{1}$} \end{footnotesize}}
\Text(38,55)[bc]{\begin{footnotesize} {$ {\rm q},\! \epsilon_{\alpha}$} \end{footnotesize}}
\Text(0,50)[c]{$\times$}
\Text(25,5)[c]{${\rm a)}$}
\SetOffset(110,0)
\DashArrowLine(0,50)(25,70){4}
\DashArrowLine(25,70)(50,90){4}
\DashArrowLine(50,10)(0,50){4}
\Vertex(25,70){2}
\Photon(0,50)(50,50){2}{7}
\Gluon(25,70)(50,70){-3}{3}
\Line(25,70)(50,70)
\BCirc(0,50){3.5}
\Text(44,76)[bc]{\begin{footnotesize} {$ {\rm g}$} \end{footnotesize}}
\Text(0,50)[c]{$\times$}
\Text(25,5)[c]{${\rm b)}$}
\SetOffset(220,0)
\DashArrowLine(0,50)(50,90){4}
\DashArrowLine(50,10)(25,30){4}
\DashArrowLine(25,30)(0,50){4}
\Vertex(25,30){2}
\Photon(0,50)(50,50){2}{7}
\Gluon(25,30)(50,30){3}{3}
\Line(25,30)(50,30)
\BCirc(0,50){3.5}
\Text(0,50)[c]{$\times$}
\Text(25,5)[c]{${\rm c)}$}
\SetOffset(330,0)
\DashArrowLine(0,50)(50,90){4}
\DashArrowLine(50,10)(0,50){4}
\Photon(0,50)(50,60){2}{7}
\Gluon(4,49.2)(50,40){-3}{6}
\Line(0,50)(50,40)
\BCirc(0,50){3.5}
\Text(0,50)[c]{$\times$}
\Text(25,5)[c]{${\rm d)}$}
\end{picture}
\end {center} 
 \caption{SCET diagrams. The cross indicates the insertion of all the leading 
and subleading effective theory 
   operators. The leading and subleading couplings of collinear quarks 
   and gluons are given, for instance, in Ref.~\cite{ps1}.} 
\end{figure}
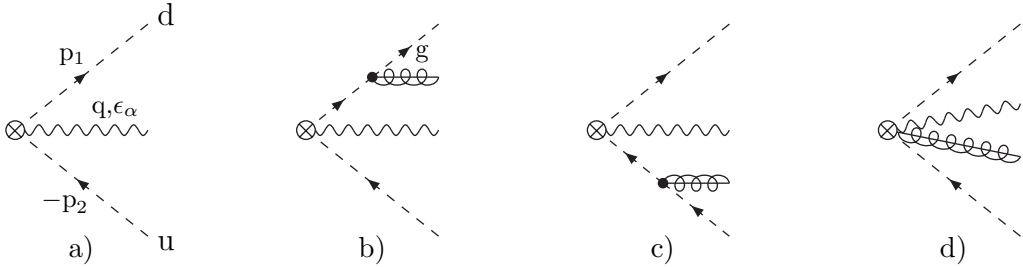 
Let us consider the matrix element of the weak current $\bar d 
\gamma^\mu P_L u$ ($P_L=(1-\gamma^5)/2$) between the vacuum and a 
state with a transverse polarised vector meson and a photon:
\bea 
\langle V(p,\eta) \gamma (q,\epsilon) |  (\bar d \gamma^\mu P_L u)(0) | 0 \rangle 
 &=& 
- i e\, \epsilon_{\alpha}^{*} \int  {\rm d}^4 x \,  e^{i q\cdot x} 
\langle V(p,\eta) | {\rm T}[ (\bar d \gamma^\mu P_L u)(0) , j^\alpha_{\rm e.m.}(x)]| 0 \rangle \nn \\ 
 & \equiv &  - i e\, \epsilon_{\alpha}^{*}  \,  \langle V(p,\eta) |  T^{\mu\alpha} | 0 \rangle\, . 
\label{weak} 
\eea 
The conservation of the electromagnetic and the
weak currents implies in the usual way the following Ward identities 
for the matrix element of the time-ordered product in
Eq.~(\ref{weak})~\cite{GrPi,KhWy}:
\bea 
q_\alpha \,   \langle V(p,\eta) |  T^{\mu\alpha} | 0 \rangle 
& = &  {i\over 2}  \, (Q_d -Q_u) f_V m_V \, \eta^{\mu *} \, , \label{emWI}\\ 
 (p_\mu + q_\mu) \,    \langle V(p,\eta) |  T^{\mu\alpha} | 0 \rangle 
 &=& {i\over 2}  \, (Q_d -Q_u) f_V m_V \, \eta^{\alpha *}  \, . 
\label{WI} 
\eea 
We checked that the electromagnetic Ward identity, Eq.~(\ref{emWI}), 
is satisfied at leading and subleading order but its check does not 
require any cancellation between the matrix elements of the various 
subleading SCET operators. We will therefore focus onto the weak 
current Ward identity, Eq.~(\ref{WI}). 
 
In the following we project the T-product (\ref{weak}) onto SCET operators, and 
then use the matrix elements computed in the previous section to 
reproduce Eq.~(\ref{WI}). For generality we keep the Dirac structure 
of the weak current completely general $\bar d \Gamma u$ and define the
operator
\bea
T^\alpha(q) = \int  {\rm d}^4 x \,  e^{i q\cdot x}
 {\rm T}[ (\bar d \Gamma u)(0) , j^\alpha_{\rm e.m.}(x)]\,.
\eea
For definiteness, we take the photon momentum along the $\bn$ direction 
$q_\mu = E_\gamma \bn_\mu$, and  the meson moving with a large momentum component 
along the $n_\mu$ direction.
Expanding the QCD graphs contributing to the $T-$product 
and keeping terms up to subleading order gives the following tree level
matching
\bea\label{Tgeneral}
T^{\alpha} &=&T^{\alpha}_{(0)} +  T^{\alpha}_{(1)}\,, \\
T^{\alpha}_{(0)} & = &
 i\frac{Q_d}{\omega_1} \bar\chi_{n,\omega_1}^{(d)} \gamma^\alpha \frac{\bnslash}{2}\Gamma
\chi_{n,\omega_2}^{(u)} + 
 i\frac{Q_u}{\omega_2} \bar\chi_{n,\omega_1}^{(d)} \Gamma \frac{\bnslash}{2}
\gamma^\alpha \chi_{n,\omega_2}^{(u)} \\
T^{\alpha}_{(1)} & = &
iQ_d\left\{ \frac{2}{n\mcdot q \omega_1} 
\left[\bar\xi_n^{(d)} (iD_\perp^\alpha)^\dagger W\right]_{\omega_1} \Gamma 
\chi_{n,\omega_2}^{(u)} + \frac{1}{\omega_1 \omega_2}
\bar\chi_{n,\omega_1}^{(d)} \gamma^\alpha \frac{\bnslash}{2}\Gamma
\left[ W^\dagger i\Dslash_\perp \frac{\bnslash}{2} \xi_n^{(u)}\right]_{\omega_2}
\right.\\
& & \qquad\qquad\left.
 -\frac{1}{n\mcdot q} \frac{\omega_3^2}{\omega_1(\omega_1-\omega_3)}
\bar\chi_{n,\omega_1}^{(d)} \gamma^\alpha \left[\frac{1}{\bn\mcdot {\cal P}}
W^\dagger i\Dslash_\perp W\right]_{\omega_3} \Gamma \chi_{n,\omega_2}^{(u)}\right\}\nn\\
&+& iQ_u \left\{ \frac{2}{n\mcdot q \omega_2} 
\bar\chi_{n,\omega_1}^{(d)} \Gamma 
[W^\dagger  iD_\perp^\alpha W \xi_n^{(u)}]_{\omega_2} + 
\frac{1}{\omega_1 \omega_2}
[\bar\xi_n^{(d)} \frac{\bnslash}{2} (i\Dslash_\perp)^\dagger W]_{\omega_1}
\Gamma \frac{\bnslash}{2}\gamma^\alpha \chi_{n,\omega_2}^{(u)}
\right.\nn\\
& & \qquad\qquad\left.
 +\frac{1}{n\mcdot q} \frac{\omega_3^2}{\omega_2(\omega_2+\omega_3)}
\bar\chi_{n,\omega_1}^{(d)} \Gamma \left[\frac{1}{\bn\mcdot {\cal P}}
W^\dagger i\Dslash_\perp W\right]_{\omega_3} \gamma^\alpha 
\chi_{n,\omega_2}^{(u)}\right\}\nn\,.
\eea
On the right-hand side, integration over $\omega_i$ is implied.

Some details about this matching computation are perhaps in order.
The two-body subleading operators (the first and third lines of $T^\alpha_{(1)}$)
can be obtained from expanding the graphs with an external $d\bar u$ quark 
pair (see Fig.~1(a)). The remaining three-body operators 
(the second and fourth lines of $T^\alpha_{(1)}$) are computed by
expanding the QCD graphs in Fig.~1(b)-(d) and subtracting the insertion of the
two-body operators computed in the previous step (Fig.~2(b)-(d)).

The case of the weak current considered in Eq.~(\ref{weak}) is obtained by
taking $\Gamma \to \gamma^\mu P_L$ and expressing the operators in
Eqs.~(\ref{Tgeneral}) in terms of the subleading collinear operators introduced
in Sec.~\ref{bilinears}.
We will assume the $\alpha$ Lorentz index to be
strictly orthogonal, corresponding to real photon emission. 
This gives
\bea 
T^{\mu\alpha} &=&T^{\mu\alpha}_{(0)} +  T^{\mu\alpha}_{(1)}\,, \\ 
T^{\mu\alpha}_{(0)} & = & 
\left[ 
i g_\perp^{\mu\alpha} \left( {Q_d \over \omega_1} + {Q_u\over \omega_2} \right) + 
\epsilon_\perp^{\mu\alpha}  \left( {Q_d \over \omega_1} - {Q_u\over \omega_2} \right) 
\right] 
{{\cal J}_A -{\cal J}_V \over 2}   \,,  \\ 
T^{\mu\alpha}_{(1)} & = & 
i {\bn^\mu \over 4} \left\{ 
Q_d {{\cal V}_1^\alpha - {\widetilde {\cal V}}_1^\alpha - {\cal A}_1^\alpha +   {\widetilde {\cal A}}_1^\alpha \over \omega_1} 
+ 
 Q_u {{\cal V}_1^\alpha + {\widetilde {\cal V}}_1^\alpha - {\cal A}_1^\alpha -   {\widetilde {\cal A}}_1^\alpha \over \omega_2} 
\right\} \nn \\ 
 & &+ i {n^\mu \over 4 E_\gamma } \left\{ 
 Q_d \left({\cal V}_2^\alpha + {\widetilde {\cal V}}_2^\alpha - {\cal A}_2^\alpha -   {\widetilde {\cal A}}_2^\alpha \right) 
 + 
  Q_u \left(-{\cal V}_2^\alpha + {\widetilde {\cal V}}_2^\alpha + {\cal A}_2^\alpha -   {\widetilde {\cal A}}_2^\alpha \right) 
 \right\}  \nn \\ 
  & & \hskip -0.75cm 
- i {n^\mu \over 4}{ \omega_3^2\over E_\gamma } \left\{ 
  Q_d {{\cal V}_3^\alpha + {\cal A}_3^\alpha - i \epsilon_\perp^{\alpha\beta} 
({\cal V}_{3\beta} + {\cal A}_{3\beta}) 
            \over \omega_1 (\omega_1-\omega_3)} - 
  Q_u {{\cal V}_3^\alpha - {\cal A}_3^\alpha - i \epsilon_\perp^{\alpha\beta} 
({\cal V}_{3\beta} - {\cal A}_{3\beta}) 
            \over \omega_2 (\omega_2+\omega_3)} 
  \right\}          , 
\eea 
After the insertion of the SCET matrix elements and the 
contraction with $p^\mu + q^\mu = {\bn\cdot p \over 2}\, n^\mu + 
E_\gamma \, \bn^\mu$, the matrix element of the leading order term 
vanishes, $ ( p_\mu + q_\mu ) \langle T^{\mu\alpha}_{(0)} \rangle = 
0$.  The subleading contribution, on the other hand, gives 
\bea 
( p_\mu + q_\mu )   \langle T^{\mu\alpha}_{(1)} \rangle 
& = & +{i\over 2} Q_d f_V m_V \eta_\perp^{\alpha *}  \, 
           \left[ {g_\perp^{(v)} (u) \over 2u} - {{g_\perp^{(a)}}^\prime (u) \over 8u} - 
                                        {F(u) - G_{Vx}^{(v)} (u)\over u} 
- {{\cal V}(\vec\alpha) +{\cal A}(\vec\alpha) \over 2 \alpha_1 (1-\alpha_2)} 
\right]      \nn \\ 
& &   -  {i\over 2} Q_u f_V m_V \eta_\perp^{\alpha *}   \, 
 \left[ {g_\perp^{(v)} (u) \over 2 \bar u} - {{g_\perp^{(a)}}^\prime (u) \over 8\bar u} + 
                                           {F(u) + G_{Vy}^{(v)} (u)\over \bar u} 
+ {{\cal V}(\vec\alpha) -{\cal A}(\vec\alpha) \over  2 \alpha_2 (1-\alpha_1)}                     \right]  \nn \\ 
 &  & \hskip -2cm 
 -{1\over 2} Q_d f_V m_V \epsilon_\perp^{\alpha \beta} \eta_{\perp\beta}^* 
            \, \left[ {g_\perp^{(v)} (u) \over 2u} - {{g_\perp^{(a)}}^\prime (u) \over 8u} - 
                                         {g_\perp^{(a)} (u) - 4 G_{Vx}^{(a)} (u)\over 4 u} 
        - {{\cal V}(\vec\alpha) +{\cal A}(\vec\alpha) \over 2 \alpha_1 (1-\alpha_2)} 
\right]      \nn \\ 
 &  &  \hskip -2cm 
        +{1\over 2} Q_u f_V m_V \epsilon_\perp^{\alpha \beta} \eta_{\perp\beta}^* 
            \, \left[ {g_\perp^{(v)} (u) \over 2\bar u} + {{g_\perp^{(a)}}^\prime (u)\over 8\bar u} - 
                                         {g_\perp^{(a)} (u) + 4 G_{Vy}^{(a)} (u)\over 4 \bar u} 
     + {{\cal V}(\vec\alpha) -{\cal A}(\vec\alpha) \over 2 \alpha_2 (1-\alpha_1)} 
\right] , 
\label{WInum} 
\eea 
where the integrations $\int_0^1 {\rm d}u$ and $\int_0^1 {\rm 
  d}\alpha_1 \int_0^{1-\alpha_1} {\rm d}\alpha_2 $, respectively, are 
understood. Using the QCD equations of motion~\cite{babr,babr2} it is 
easy to verify that the $ \epsilon_\perp^{\alpha \beta} 
\eta_{\perp\beta}^*$ pieces vanish while the integral over the 
$\eta_\perp^{\alpha *}$ terms reproduces exactly the Ward identity, 
Eq.~(\ref{WI}). In proving this cancellation, the following identities
are useful
\begin{eqnarray}
& &\int_0^1 {\rm d}u \frac{1}{u\bar u} G_{Vx}^{(v)}(u) - 
\int_0^1 {\rm d}\alpha_1 \int_0^{1-\alpha_1} {\rm d}\alpha_2
\frac{{\cal V}(\alpha_i)}{\alpha_1 (1-\alpha_2)} = 0\,,\\
& &\int_0^1 {\rm d}u \frac{\bar u-u}{u\bar u} G_{Vx}^{(a)}(u) - 
\int_0^1 {\rm d}\alpha_1 \int_0^{1-\alpha_1} {\rm d}\alpha_2
\frac{{\cal A}(\alpha_i)}{\alpha_1 (1-\alpha_2)} = 0\,.\nn
\end{eqnarray}

\begin{figure}[t] 
\begin{center} 
\includegraphics[height=6cm]{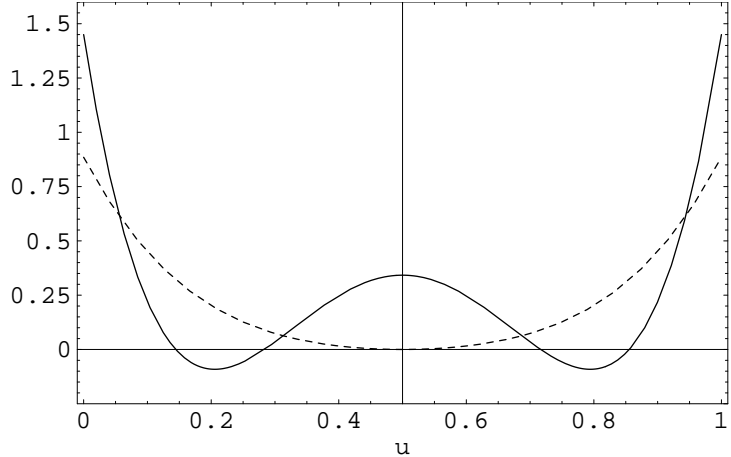} 
\end{center} 
\caption{Matrix element (modulo a factor $f_V m_V \eta_\perp^{\mu *}$) 
  of the subleading operator ${\cal V}_2^\mu$ with (solid line) and
  without (dashed line) the inclusion of the three-particle
  distribution amplitudes. Note that in the latter case we
  consistently set to zero all the twist-3 three-particle wave
  functions.  The parameterization of the wave functions as well as the
  numerical value of the various parameters correspond to the $\rho$
  meson and are taken from Ref.~\cite{babrhand,babr2} (see also
  App.~\ref{wf}).}
\label{fig:v2}
\end{figure} 

Adopting the parameterization of the vector meson wave functions given 
in Refs.~\cite{babr,babr2}, we can get an idea of the numerical impact 
of the three-particle contributions. First of all, note that the 
equations of motion of QCD allow to express the twist-3 two-particle 
wave functions in terms of the twist-2 two-particle and twist-3 
three-particle ones. This implies that, in the limit in which we 
consistently set to zero the functions $\cal V$ and $\cal A$, the Ward 
identity will be still satisfied. This is a peculiar property of the 
check that we are performing and it is not a general feature of 
exclusive decay amplitudes.  From the inspection of Eq.~(\ref{WInum}) 
we see that three-particle light-cone wave functions can enter the 
amplitude either by modifying the SCET matrix elements of the 
subleading two-particle operators ({\it e.g.} ${\cal V}_2^\mu$, ${\cal 
  A}_2^\mu$, ...)  or with new genuine contributions ({\it e.g.} 
${\cal V}_3^\mu$, ${\cal A}_3^\mu$, ...).  In our case both 
contributions are of order $10\%$ and cancel each other. We expect, 
therefore, effects of order $20\%$ in generic exclusive amplitudes 
that start at subleading order in $\lambda$. This is quite common for 
decays with vector mesons in the final state because, in this case, 
the usual $V-A$ weak current has no leading projections onto states 
with a transverse polarised meson. For pseudo-scalar mesons, on the 
other hand, we do not expect large contributions because the leading 
SCET operators have non vanishing projection and are expected to give 
the dominant contribution. 
 
Finally, note that the importance of these contributions strongly 
depends on the form of the hard scattering kernel. Consider, for 
instance, the operator ${\cal V}_2^\mu$. We plot in Fig.~\ref{fig:v2} 
its matrix element, Eq.~(\ref{v2}), with and without the inclusion of 
the three-particle contributions. In the former case, the shape of the 
distribution amplitude is changed and configurations closer to the 
center-point ($u\sim 0.5 $) become more important.

\section{Conclusions}

In this paper we classify the leading and subleading SCET operators
(in powers of $\lambda$) necessary for the analysis of exclusive
processes that involve fast moving light mesons. Some aspects of SCET
at subleading order still have to be fully understood, but, in order
to perform our analysis, one only needs the field content of the
theory and the corresponding gauge transformations. At subleading
order, both two-particle ($\sim \bar \xi \xi$) and three-particle
($\sim \bar\xi A_\perp^\mu \xi$) operators are present.

We show how to express the matrix elements of all the SCET operators
between the vacuum and a meson state in terms of the standard two- and
three-particle meson light-cone distribution amplitudes. Exact
operator identities between various SCET operators result in
corresponding identities between different distribution amplitudes.
We checked that all these relations are exactly satisfied by the
application of the QCD equations of motion. In the limit in which the
contribution of the three-particle operators is neglected, we recover
several Wandzura--Wilczek relations between wave functions of
geometrical twist-2 and twist-3. This reflects the fact that operators
with a given $\lambda$ scaling have fixed dynamical twist ({\it i.e.}
they have a fixed $\Lambda/Q$ suppression); due to the mismatch
between dynamical and geometric twist, their matrix elements contain
contributions from wave functions of different geometrical twist.

The subleading collinear structures considered here appear as building blocks
for SCET operators contributing to power suppressed processes.
We showed that Lorentz invariance (manifested in the SCET as reparameterization
invariance) severely constrains the Wilson coefficients of such subleading
operators, and connects them to the coefficients of the leading operators.

In order to further validate our results and get a feeling for the
size of the three-particle contributions, we checked a Ward identity
for the final state emission of a photon from a transverse polarized
vector meson. Comparing the exact result with the computation in the
Wandzura-Wilczek limit (in which the three-particle distribution
amplitudes are set to zero), we find that corrections of order
$O(20\%)$ at the amplitude level are possible.

In general three-particle light-cone wave functions are expected to be
more important in processes involving vector mesons in the final state
({\it e.g.} $B\to K^* \pi$, ...). In fact, the amplitude for the
production of a transverse polarised vector meson does not receive
twist-2 contributions (in SCET language, the matrix elements of the
leading order operator are vanishing). On the contrary, decays that
involve only pseudo-scalar mesons ({\it e.g.} $B\to \pi\pi$, ...)
start at twist-2 and, therefore, three-particle corrections are small.

\section*{ACKNOWLEDGEMENTS} 
We thank Martin Beneke and Thorsten Feldmann for interesting
discussions. D.P. is grateful to Michael Gronau and Iain Stewart for
useful discussions.  This work is partially supported by the Swiss
National Fonds and by the DOE under Grant No. DOE-FG03-97ER40546 and
by the US National Science Foundation Grant PHY-9970781.

\appendix 
\section{Notation and conventions} 
\label{notation} 

We use the standard convention for the metric tensor in Minkowski
space $ g^{\mu \nu}={\rm diag}(1,-1,-1,-1)$ and the totally
antisymmetric tensor $ \varepsilon^{\mu\nu\rho\sigma}$ is defined with
$\varepsilon_{0123}=-\varepsilon^{0123}=1$.  For the Dirac matrices we
use the standard conventions, in particular
$\gamma^{5}=i\gamma^{0}\gamma^{1}\gamma^{2}\gamma^{3}$ and define
$\sigma_{\mu\nu}$ by 
\bea 
\sigma_{\mu\nu}\equiv\frac{i}{2} \left[\gamma_{\mu},\gamma_{\nu}\right]\,. 
\eea 
From this we get the relation $ \sigma_{\mu\nu}\gamma^{5}=-\frac{i}{2} 
\varepsilon_{\mu\nu\rho\sigma}\sigma^{\rho\sigma}$, and its inverse
$\sigma_{\alpha\beta} = -\frac{i}{2} \varepsilon_{\alpha\beta\mu\nu}
\sigma^{\mu\nu}\gamma_5$.
The dual gluon field strength tensor is defined by 
$\widetilde{G}_{\mu\nu}=\frac{1}{2}\varepsilon_{\mu\nu\rho\sigma}G^{\rho\sigma}$. 

The hadrons are taken to be moving with large momentum along the 
light-cone direction $n$. The opposite direction light cone
vector $\bar{n}$ is chosen such that $n^{2}=\bar{n}^{2}=0$ and $n 
\cdot \bar{n}=2$. Momenta are decomposed along the light cone as 
$p=(p_-,p^+,p_{\perp})$ with
\bea
p^{\mu}=\frac{\bar{n} \mcdot p}{2} n^{\mu}+\frac{n \mcdot p}{2}\bar{n}^{\mu}+
p_{\perp}^{\mu} 
\,.
\eea 
We thus define the perpendicular component of the $\gamma$ matrices by 
\bea 
\gamma_{\perp}^{\mu}\equiv
\gamma^{\mu} - \frac{\nslash}{2}\bar{n}^{\mu}-\frac{\bnslash}{2}n^{\mu} \,.\nn
\eea
Similarly we write the perpendicular component of the metric tensor:
\bea
g_{\perp}^{\mu \nu} \equiv 
g^{\mu\nu}- \frac12 n^{\mu}\bar{n}^{\nu} - \frac12 n^{\nu} \bar{n}^{\mu}\,. 
\eea
We define the perpendicular components of the total antisymmetric tensor as
\bea
\varepsilon_{\perp}^{\mu\nu} \equiv 
\frac{1}{2} \varepsilon^{\mu\nu\rho\sigma}\bar{n}_{\rho}n_{\sigma}\,.
\eea
It satisfies the following relations which are useful in practical calculations
\bea\label{id1}
\varepsilon_{\perp}^{\mu\alpha} \varepsilon^\perp_{\alpha\beta} g_\perp^{\beta\nu}  
 &=& - g_{\perp}^{\mu \nu}\, ,   \\ 
\label{id2}
\frac{\bnslash}{2}\gamma_{\perp}^{\mu} \gamma_{\perp}^{\nu} &=& \frac{\bnslash}{2} 
g_{\perp}^{\mu \nu}-i \varepsilon_{\perp}^{\mu\nu} 
\frac{\bnslash}{2} \gamma^{5} \, ,   \\ 
\label{id3}
\varepsilon_{\perp}^{\mu\nu}p_{\perp\mu}p_{\perp\nu}n^{\rho} &=& 
\varepsilon^{\mu\nu\rho\sigma} 
p_{\perp\mu}p_{\perp\nu}n_{\sigma}\, ,  \\
\label{id4}
\frac{\bnslash}{2} \gamma_\mu^\perp \gamma_5 &=& i\varepsilon^\perp_{\mu\nu}
\frac{\bnslash}{2}  \gamma_\nu^\perp\,.
\eea

\section{Wavefunction parameterizations}
\label{wf}

We collect in this Appendix simple parameterizations of the light-cone
wave functions used in the numerical evaluations in the text,
following \cite{babrhand,babr2}.  For more details about their
derivation and the computation of the nonperturbative parameters
appearing in these formulas, we refer to
\cite{ChZh,Gorsky,fily1,fily2,babr,babrhand,babr2,pb,conformal}.

We start by listing the pseudoscalar wave functions
$\phi_p(u),\phi_\sigma(u), \phi_{3\pi}(\alpha_1,\alpha_2,\alpha_3)$.
Working at next-to-leading order in the conformal expansion, they can
be written as (with $\xi = 2u-1$)
\bea
\phi_p(u) &=& 1 + 15 R [3\xi^2 - 1]\,, \\
\phi_\sigma(u) &=& 6u (1-u) \left\{ 1 + (5-\frac12 \omega_{1,0}) R 
[\frac{15}{2} \xi^2 - \frac32]\right\}\,, \\
\phi_{3\pi}(\alpha_i) &=& 360 \alpha_1 \alpha_2 \alpha_3^2 \left( 1 +
\frac12 \omega_{1,0} (7\alpha_3-3)\right)\,. 
\eea

For the vector meson case, we give only the wave
functions corresponding to the $\rho$ meson.  Working again at
next-to-leading order in the conformal expansion, the twist-2
two-parton wave functions are
\bea
\phi_{\parallel,\perp}(u) &=& 6u\bar u
\left[ 1 + a_2^{\parallel,\perp} \frac32 [5(2u-1)^2 - 1] \right]\,.
\eea
The remaining twist-3 two-parton wave functions $g_\perp^{(v,a)}(u)$
and $h_\parallel^{(s,t)}(u)$
can be expressed with the help of the equation of motion in terms of
the twist-3 three-parton wave functions. The leading terms in their
conformal expansions are
\bea
\{ {\cal V}, {\cal T}\}(\alpha_1,\alpha_2,\alpha_3) &=&  
540 \zeta_3 \omega_3^{V,T} (\alpha_1 - \alpha_2) \alpha_1 \alpha_2 \alpha_3^2\,, \\
{\cal A}(\alpha_1,\alpha_2,\alpha_3) &=& 360 \zeta_3 
\alpha_1 \alpha_2 \alpha_3^2 [1 + \omega_3^A
\frac12 (7\alpha_3 - 3)]\,.
\eea
We follow here the notations of \cite{babrhand,babr2}. The numerical
values of the nonperturbative constants are $a_2^\parallel = 0.18\pm 0.10,
a_2^\perp = 0.2\pm 0.1, \zeta_3=0.032, \omega_3^V=3.8, \omega_3^T = 7.0,
\omega_3^A=-2.1$ (at the scale $\mu^2 = 1$ GeV$^2$) \cite{babrhand}. 
Using the equations of motion, the chiral-even two-particle twist-3
wave functions are determined as
\bea
g_\perp^{(a)}(u) &=& 6u\bar u\left[
1 + \left\{ \frac14 a_2^\parallel + \frac53 \zeta_3 [1 - \frac{3}{16}\omega_3^A]
+ \frac{15}{16} \zeta_3 \omega_3^V \right\}(5\xi^2-1)\right]\,, \\
g_\perp^{(v)}(u) &=& \frac34 (1 + \xi^2) + 
\left( \frac37 a_2^\parallel + 5\zeta_3 \right) 
(3\xi^2-1)\\
&+& \left[ \frac{9}{112} a_2^\parallel + \frac{15}{64} \zeta_3 (3\omega_3^V -
\omega_3^A) \right] (3-30 \xi^2 + 35 \xi^4)\nn\,.
\eea
The corresponding results for the chiral-odd two-particle twist-3
wave functions are
\bea
h_\parallel^{(t)}(u) &=& 3\xi^2 + \frac32 a_2^\perp \xi^2 (5\xi^2-3) +
\frac{15}{16} \zeta_3 \omega_3^T (3-30 \xi^2 + 35 \xi^4)\\
h_\parallel^{(s)}(u) &=& 6u \bar u \left\{ 
1 + \left( \frac14 a_2^\perp + \frac58 \zeta_3 \omega_3^T \right)
(5\xi^2-1)\right\}\,.
\eea

\end{document}